# *Long-term decrease in colouration: a consequence of climate change?*

David López-Idiáquez[1,2*]; Céline Teplitsky[1]; Arnaud Grégoire; Amélie Fargevieille[1,3]; María Del Rey[1]; Christophe de Franceschi[1]; Anne Charmantier[1]; Claire Doutrelant[1]

[1]CEFE, CNRS, Univ Montpellier, EPHE, IRD, Montpellier, France.
[2]Department of Plant Biology and Ecology, University of the Basque Country (UPV/EHU), Leioa, 48940, Spain.
[3]Department of Biological Sciences, College of Sciences and Mathematics, Auburn University, Auburn AL, 36849, USA.

*Corresponding author: davididiaquez@gmail.com

**Abstract**

Climate change has been shown to affect fitness-related traits in a wide range of taxa; for instance, warming leads to phenological advancements in many plant and animal species. The influence of climate change on social and secondary sexual traits, that are associated with fitness due to their role as quality signals, is however unknown. Here, we use more than 5800 observations collected on two Mediterranean blue tit subspecies (*Cyanistes caeruleus caeruleus* and *C.c. ogliastrae*) to explore whether blue crown and yellow breast patch colourations have changed over the past 15 years. Our data suggests that colouration has become duller and less chromatic in both sexes. In addition, in the Corsican *C.c. ogliastrae,* but not in the mainland *C.c. caeruleus*, the decrease is associated with an increase in temperature at moult. Quantitative genetic analyses do not reveal any microevolutionary change in the colour traits along the study period, strongly suggesting that the observed change over time was caused by a plastic response to the environmental conditions. Overall, this study suggests that ornamental colourations could become less conspicuous due to warming, revealing climate change effects on sexual and social ornaments and calling for further research on the proximate mechanisms behind these effects.

**Translated abstract (French):**

Il a été montré que les changements climatiques affectent les traits de reproduction de nombreux taxons ; par exemple, le réchauffement climatique entraîne des avancements phénologiques. Qu'en est-il des traits sociaux et sexuels secondaires? Ici, nous avons analysé plus de 5800 observations collectées sur deux sous-espèces de mésanges bleues méditerranéennes (*Cyanistes caeruleus caeruleus* et *C. c. ogliastrae*) pour explorer si les colorations de la couronne bleue et des taches jaunes de la poitrine ont changé au cours des 15 dernières années. Nos données suggèrent que la coloration est devenue plus terne et moins chromatique chez les deux sexes. De plus, pour la sous-espèce corse, cette diminution est associée à une augmentation de la température au moment de la mue. Les analyses génétiques quantitatives réalisées ne révèlent aucun changement microévolutif dans les traits de couleur au cours de la période d'étude, ce qui suggère fortement que le changement observé au fil du temps a été causé par une réponse plastique aux conditions environnementales. Dans l'ensemble, cette étude suggère que les colorations ornementales pourraient devenir moins visibles en raison du réchauffement climatique et appelle à des recherches supplémentaires sur les mécanismes pouvant créer cet effet.

## Introduction

Environmental conditions play a key role in modulating the life cycles of organisms (Stenseth et al. 2002). As a consequence, in many taxa, trait expression changes as a result of environmental heterogeneity, especially in relation to current rapid climate change (plants: Peñuelas et al. 2002; McDowell et al. 2020, fish: Crozier and Hutchings 2014; Asch et al. 2019, mammals: Réale et al. 2003; Boutin and Lane 2014 and insects: Pureswaran et al. 2018; Kellermann and van Heerwaarden 2019). One of the most commonly reported responses regards plant and animal phenology (Parmesan 2007; Hällfors et al. 2020) whereby many species are advancing their breeding periods as a consequence of global warming (Hällfors et al. 2020). By contrast, less attention has been given to the impacts of climate change on other fitness-related traits, such as social or secondary sexual traits.

Ornaments have evolved to signal individual quality to mates or competitors determining mating opportunities and/or the outcome of intra-sexual interactions (Andersson 1994) however, whether climate change influences the expression and contemporary evolution of ornaments is unclear (Svensson 2019). Yet, a reduction in ornament intensity and variance can reduce their signalling potential and the strength of sexual selection acting on them (Cockburn et al. 2008), something that can compromise the adaptive capabilities of a population (Gómez-Llano et al. 2021).

An ornament's reliability is often associated with its cost and/or condition-dependence (Pomiankowski 1987; Cotton et al. 2004) and as a consequence, ornament expression is expected to be sensitive to variation in environmental conditions (Breckels and Neff 2013; López-Idiáquez et al. 2016*a*). In male guppies (*Poecilia reticulata*), for instance, experiments have revealed that their orange ornament hue is highest for a water temperature of 28 ºC while lower at colder and warmer temperatures (Breckels and Neff 2013). Further, the environment can also modulate the relative costs and benefits associated with ornament expression (Vergara et al. 2012). For example, in lions (*Panthera leo*) mane length and darkness, which are secondary sexual traits, have temperature-dependent costs as the individuals with longer and darker manes suffer from reduced sperm quality and lower ability to obtain food at high temperatures (West and Packer 2002; Patterson et al. 2006). Thus, given the current rapid global warming, the costs associated with the production and maintenance of ornaments could be expected to change and, at least in warmer areas, climate change could constrain the expression of secondary sexual traits.

In birds and in other taxa, the expression of conspicuous coloured patches is usually driven by sexual and/or social selection (Stuart-Fox and Ord 2004; Dale et al. 2015). Plumage colouration can be produced by different mechanisms: the deposition of pigments such as carotenoids or melanin, the microstructure of the feather or by a combination of both (Hill and McGraw 2006). Due to their different developmental mechanisms, each type of colouration may respond differently to environmental variation. On the one hand, carotenoid-based traits are linked to factors such as diet, stress, or parasite prevalence and their signalling role is traded-off with anti-oxidant and immune functions thus, they are expected to be highly sensitive to the environmental variation (Hill et al. 2002; Mougeot et al. 2010). On the other hand, structural colourations are less costly to produce (Prum et al. 2009) and their expression constraints may be more indirectly linked to the environment (e.g. feather development time; Griggio et al. 2009), thus they are expected to be less reactive to environmental variation (Prum 2006; Janas et al. 2020). However, all colourations present some degree of environmental dependence, and can be driven by the inter-annual variation in climatic conditions (e.g. for birds; melanin: Jensen et al. 2006, carotenoids: Reudink et al. 2015, structural: Masello et al. 2008). In general, we can expect colouration to become duller and less chromatic due to climate change, since warmer temperatures have been negatively associated with body condition (McLean et al. 2018) and increased infection risk by parasites like haemosporidians (Garamszegi 2011) that can negatively affect colouration (del Cerro et al. 2009). However, it is also important to consider that the strength and direction of the consequences of climate change on ornamental colourations may be different between colours and populations, as climate change effects differ across geographic areas (IPCC 2018).

To the best of our knowledge, only four long term studies have explored the potential influence of climate change on the expression of ornamental colourations, two focusing on the size of an achromatic patch in migratory bird species (Scordato et al. 2012; Evans and Gustafsson 2017), one on the intensity of the carotenoid-based colouration in a sedentary bird species (Laczi et al. 2020) and another on the degree of melanisation in dragonflies (Moore et al. 2021). The first found no variation on the achromatic white-wing bar size of Hume's warblers in the Himalayas (*Phylloscopus humei*) along 25-years, despite a mean 2ºC increase in the study area (Scordato et al. 2012). Two found negative associations with climate change. The size of the achromatic white forehead patch size in collared flycatchers (*Ficedula albicollis*) breeding in Sweden decreased, hypothetically, as a consequence of the estimated 1.5ºC rise along a 34-year period (Evans and Gustafsson 2017). Also, warming led to a decrease in the degree of melanisation in ten dragonfly species across North America between 2005 and 2019 (Moore et al. 2021). The fourth study, on the contrary, described an increase in the yellow chest colouration over eight years in a Hungarian great tit (*Parus major*) population due to the increase in temperature and decrease in precipitations (Laczi et al. 2020). To date, nothing is known on how climate change affects the intensity of other types of colourations such as structural colours which, despite being understudied, play an important role as condition-dependent signals in different taxa such as birds, reptiles or arachnids (White 2020). Further, from the aforementioned studies only one reported a microevolutionary response (using estimated breeding values) of the ornamental colours to the environmental change (Evans and Gustafsson 2017). Still, identifying whether changes at the phenotypic level occur through microevolution or are the product of phenotypic plasticity is crucial to understand the consequences of climate change, as only microevolution can ensure the adaptation to continued environmental change (van Buskirk and Steiner 2009; Duputié et al. 2015).

Here, we explored the effects of climate change on two colouration types of male and female blue tits (*Cyanistes caeruleus*). We used 15 years of data collected in two subspecies, *C. c. caeruleus* on the French Mediterranean mainland and *C.c. ogliastrae* in Corsica near the south edge of the species distribution. Specifically, we studied the UV-blue colouration of the crown (structural colouration) and the yellow colouration of the breast patch (carotenoid-based colouration), both renewed annually during moult in summer. These two colourations are condition-dependent in both sexes in our populations (Doutrelant et al. 2012) and may play a role as signals in social and sexual contexts (Hunt et al. 1999; Sheldon et al. 1999; Alonso-Alvarez et al. 2004; Limbourg et al. 2013; Midamegbe et al. 2013; Doutrelant et al. 2020, but see Parker 2013). Our specific objectives were to determine whether: i) blue tit ornamental colourations showed a temporal trend in their expression, ii) temperature and precipitations have changed along the study period, iii) the temporal variation observed in trait expression was explained by the variation in temperature and precipitation and iv) the change shown at the phenotypic level was due to microevolution, by exploring the temporal trends in trait breeding values. Given the expected negative impacts of warming and droughts on bird body condition and health for species living in hot and dry conditions (Garamszegi 2011; McLean et al. 2018; McKechnie and Wolf 2019), we predicted that if temperature is increasing and/or precipitation is decreasing in the study area, we would find a decline in blue tit ornamental colourations due to their condition-dependence (Doutrelant et al. 2012). In addition, given that it has been shown that blue tit colouration is heritable in both our (Charmantier et al. 2017) and in other populations (Hadfield et al. 2006; Drobniak et al. 2013), we evaluated the extent to which a genetic change contributes to the change described at the phenotypic level.

## Methods

### Study area and general procedures

This study was conducted between 2005 and 2019 in two Mediterranean areas equipped with nest-boxes. The first area is located in La Rouvière forest (D-Rouvière) in the vicinity of Montpellier (subspecies *caeruleus*). The second is located on the island of Corsica (subspecies *ogliastrae*) and includes three different study sites in Northwest Corsica (D-Muro, E-Muro and E-Pirio; Charmantier et al. 2016). Blue tits were captured when nestlings were nine days old. At that time, each bird was ringed with a uniquely numbered metal ring and six blue feathers from the crown and eight yellow feathers from the chest were collected to assess the colouration using spectrophotometry in the lab. The captures and the sampling of the feathers were done following a standardized protocol, mostly by permanent staff and PhD students (88% of the captures). We computed chromatic and achromatic colour variables using Avicol v2 or the R package "*pavo*" (Gomez 2006; Maia et al. 2019). Specifically, for the blue crown colouration we computed UV chroma and brightness and for the yellow breast patch colouration we computed yellow chroma and brightness, following precedent studies (Andersson et al. 1998; Doutrelant et al. 2008, 2012; Fargevieille et al. 2017; see Supplementary Material (SM)-1 for further details on how the feathers were sampled and measured and how colour was extracted).

Blue tits undergo one moult per year. First, the post-juvenile moult that is limited to the head, body and a variable number of flight feathers taking place in summer approximately at two months of age (Shirihai and Svensson 2018; Stenning 2018). Then, as adults, the complete post-breeding moult in which blue tits renew all their feathers mainly between mid-June and mid-September (Shirihai and Svensson 2018).

**Climatic variables**

Daily values of mean temperature (i.e., average between the maximum and minimum temperature of the day) and precipitation between 2004 and 2018 were obtained from the French national meteorological service. Following Bonamour et al. (2019), we used weather information from the stations of Saint Martin de Londres on the mainland (about 24 km from D-Rouvière) and Calvi in Corsica (9-19 km to the Corsican study sites). Temperature from the meteorological stations was highly correlated with the local temperature at each study site (see Bonamour et al. 2019 for further information). With this information, we computed two variables for each population: the average temperature during moult (average temperature between 1$^{st}$ June to 30$^{th}$ September in the previous year) and average precipitation during moult (i.e. across the same period). The time interval between June and September in the previous year was selected to capture the climatic conditions experienced by the blue tits during moulting.

**Statistical analyses**

(i) Temporal trends in colouration

To test for a change in the coloured traits over time, we fitted a series of Linear Mixed Models (LMMs) with a normal distribution of errors, one for each of the four colour components in the two subspecies separately. These models included the coloured traits as dependent variables and year (as a continuous variable) as an explanatory term scaled to mean of zero and standard deviation of unity. In addition, sex and its interaction with year were included as covariates. Individual identity, year (as a categorical variable) and site (in the Corsican models) were included as random factors to account for the non-independence of data within sites, years and individuals.

(ii) Temporal trends in climatic variables

The temporal trends in the climatic proxies (average temperature and precipitation during moult), for the period 2004-2018, were analysed separately for each population by fitting a Linear Model (LM) with a normal distribution of errors that included the climatic indices as dependent variables and year (as a continuous variable) as an explanatory term.

(iii) Association between the coloured traits and the climatic variables

We fitted LMMs with a normal distribution of errors to analyse the association between the colourations and the climatic proxies in the two subspecies separately. We included the coloured traits as dependent variables in separate models and average temperature and precipitation during moult as explanatory terms. Sex and its interaction with the climatic variables were also included to control for possible sex-dependent responses. Year (as a continuous variable) was included as a covariate in the D-Rouvière models but not in the Corsican models as it was collinear with the climatic variables (VIF>3, Zuur et al. 2010; see Table S2.1). All models included year (as a categorical variable), individual identity and site (in the Corsican models) as random factors. Predictors were scaled to mean of zero and standard deviation of unity.

(iv) Quantitative genetics analyses

In each population, we fitted a univariate animal model with gaussian errors for each coloured trait and sex to estimate the heritability and predicted individual breeding values for each trait.

As fixed effects, we included year (as a continuous variable) and site (in the Corsican models). The random effects decomposed the phenotypic variance ($V_P$) into four components, namely: additive genetic variance ($V_A$), permanent environment variance ($V_{PE}$, estimated using the repeated observations of the individuals in different years), variance associated with measurement year ($V_{YR}$) and residual variance ($V_R$). The $V_A$ was estimated by incorporating a relatedness matrix based on a social pedigree of our population. Extra-pair paternity occurs in our populations (average of 18.4% extra-pair youngs found between 2000-2003; Charmantier and Blondel 2003), which may lead to heritability underestimated by up to 17% (Charmantier and Réale 2005). For those individuals with unidentified parents a dummy code was assigned to represent the missing parent so that the sibship information was retained. We used the "*prunePed*" function of the MCMCglmm package (Hadfield 2010) to retain only the informative individuals in each analysis. The different pruned pedigrees used in the analysis had between 1129 and 1864 observations and a maximum pedigree depth of 16 generations (see SM-3 for further information on the pedigrees used). These models were run with the MCMCglmm package with a total of 2.500.000 iterations, including a burn-in period of 500.000 iterations and a thinning interval of 2.000 iterations. For random effects we used parameter expanded priors (V=1, nu=1, alpha.mu=0, alpha.V=1000), we also fitted additional models using alternative priors that showed that the quantitative genetic estimates were robust (for further information see SM-4). Calculations of the quantitative genetic estimates were done on the MCMC posterior distributions in order to propagate the uncertainty in parameter estimates (Evans and Gustafsson 2017; Bonnet et al. 2019). In all models, autocorrelation values were lower than 0.1 and effective sample size was at least 1000. We also checked that models satisfied convergence criteria based on the Heidelberg and Welsh convergence diagnostic (Heidelberger and Welch 1981). To estimate the temporal trends in breeding values we fitted for each trait a linear regression of the Best Linear Unbiased Predictors (BLUPs) for the additive genetic individual effect, against the individual mean value of its breeding years as previously done in Bonnet et al. (2019). To account for the uncertainty around BLUP estimations (Hadfield et al. 2010), we used their full posterior distribution to estimate the time trend. For each of the 1000 interactions, we run the regression of BLUP's against the individual mean values of the breeding year and then the overall time trend was estimated by the posterior distribution of these 1000 slopes. We used individual mean value of year rather than hatching year as it reflects when the individual was contributing to the temporal trends as a breeder and because it allows including individuals for which hatching year is unknown. Using a less conservative approach (namely, not including year as continuous fixed effect) yield similar results (see Table S5.1). Maternal effects were not considered in the animal models, something that led to a slight overestimation (mean 0.011, range: 0.004-0.027) of the heritabilities (for further details see SM-6).

Mixed models were run with the packages *lmerTest* (Kuznetsova et al. 2017) or *MCMCglmm* (Hadfield 2010) in R (Version 3.6.3; R Core Team 2019). We computed the conditional $R^2$, that represents the percentage of total variance explained by both the fixed and random effects ($R^2_{cond}$), and the marginal $R^2$, that represents the percentage of total variance explained by the fixed effects only ($R^2_{mar}$; Nakagawa and Schielzeth 2013), using the package *MuMIn* (Barton 2019). Finally, Variance Inflation Factors (VIFs) were estimated using the package *usdm* (Babak 2015) to estimate the collinearity between the explanatory variables. Including age (1-year-old vs ≥2-year-old) did not change the obtained results (see Tables S7.1-S7.4). Data used in this study have been deposited in the Dryad Digital Depository (https://doi.org/10.5061/dryad.w6m905qr9; López-Idiáquez et al. 2022).

## Results

### Coloration across years

**Blue crown:** In both Corsica and D-Rouvière, we found a significant year by sex interaction showing a decrease in the UV chroma with time (Table 1, Fig. 1) that was slightly stronger in males (Corsica: -0.019±0.005, $F_{1,13.05}$=14.63, p=0.002; D-Rouvière: -0.012±0.005, $F_{1,13.00}$=5.238, p=0.039) than in females (Corsica: -0.015±0.004, $F_{1, 13.07}$=10.05, p=0.007; D-Rouvière: -0.008±0.004, $F_{1,13.17}$=4.528, p=0.052). In both models, the fixed effects explained approximately 35% of the variance ($R^2_{mar}$) and around 70% was explained by the combined fixed and random effects ($R^2_{cond}$; Table 1). Regarding brightness, while no significant change over time was found in Corsica, it significantly decreased in both sexes in D-Rouvière (Table 1, Fig. 1; $R^2_{mar}$: 0.148, $R^2_{cond}$: 0.502).

**Yellow breast patch:** In Corsica, we found a significant year by sex interaction ($R^2_{mar}$=0.145, $R^2_{cond}$:0.368; Table 1, Fig.1) showing that chroma decreased with time with a stronger effect in males -0.052±0.012, $F_{1, 13.29}$=17.57, p=0.001) than in females (-0.030±0.012, $F_{1, 13.07}$= 6.213, p=0.026). In D-Rouvière, a marginal negative association was found between chroma and year (p=0.087). Regarding brightness, in Corsica a significant interaction between year and sex was found (Table 1, Fig,1, $R^2_{mar}$: 0.199, $R^2_{cond}$: 0.391). In both sexes, brightness decreased with year (males: -1.288±0.320, $F_{1,13.17}$=16.175, p=0.001; females: -1.597±0.310, $F_{1,13.20}$=26.497, p<0.001). In D-Rouvière brightness decreased with time with no interaction between year and sex ($R^2_{mar}$: 0.169, $R^2_{cond}$: 0.507; Table 1, Fig. 1; see Tables S8.1 and S8.3 for further details).

**Table 1**. Linear temporal trends for blue crown and yellow breast patch colourations of the blue tits in Corsica and D-Rouvière. Significant ($p<0.05$) variables are in bold. $R^2_{cond}$ represents the variance explained by both the fixed and random effects included in the model, $R^2_{mar}$ represents the variance explained by the fixed factors alone.

| | **Corsica** | | | | **D-Rouvière** | | | |
|---|---|---|---|---|---|---|---|---|
| **Blue crown UV chroma (n. obs.: Corsica=3867; D-Rouvière=2068)** | | | | | | | | |
| **Fixed effects** | Est | SE | F | P | Est | SE | F | P |
| Year | **-0.019** | **0.004** | **$F_{1,13.03}$=12.524** | **0.003** | **-0.011** | **0.004** | **$F_{1,13.03}$=5.038** | **0.042** |
| Sex(fem) | **-0.035** | **0.0008** | **$F_{1,2213.7}$=1614.7** | **<0.001** | **-0.038** | **0.001** | **$F_{1,1169.4}$=1024.9** | **<0.001** |
| Year*sex(fem) | **0.004** | **0.0008** | **$F_{1,2790.5}$=23.383** | **<0.001** | **0.002** | **0.001** | **$F_{1,1407.8}$=3.927** | **0.047** |
| | $R^2_{cond}$:0.740; $R^2_{mar}$: 0.365 | | | | $R^2_{cond}$:0.696; $R^2_{mar}$: 0.338 | | | |
| **Blue crown brightness (n. obs.: Corsica=3867; D-Rouvière=2068)** | | | | | | | | |
| **Fixed effects** | Est | SE | F | P | Est | SE | F | P |
| Year | -0.756 | 0.586 | $F_{1,13.06}$=1.287 | 0.276 | **-1.557** | **0.701** | **$F_{1,13.01}$=5.670** | **0.033** |
| Sex (fem) | **-2.378** | **0.118** | **$F_{1, 2125.6}$=399.71** | **<0.001** | **-2.332** | **0.189** | **$F_{1,1097.0}$=150.75** | **<0.001** |
| Year*sex(fem) | 0.190 | 0.117 | $F_{1,2774.0}$=2.653 | 0.103 | -0.195 | 0.187 | $F_{1,1337.7}$=1.093 | 0.295 |
| | $R^2_{cond}$:0.431; $R^2_{mar}$: 0.094 | | | | $R^2_{cond}$: 0.502; $R^2_{mar}$: 0.148 | | | |
| **Yellow breast patch chroma (n. obs.: Corsica=3890; D-Rouvière=2007)** | | | | | | | | |
| **Fixed effects** | Est | SE | F | P | Est | SE | F | P |
| Year | **-0.051** | **0.018** | **$F_{1,13.15}$=12.673** | **0.003** | -0.040 | 0.021 | $F_{1,13.07}$=3.412 | 0.087 |
| Sex (fem) | **-0.068** | **0.004** | **$F_{1,2074.8}$=277.11** | **<0.001** | **-0.012** | **0.006** | **$F_{1,1015.5}$=4.100** | **0.043** |
| Year*sex(fem) | **0.020** | **0.004** | **$F_{1,2731.1}$=25.071** | **<0.001** | 0.001 | 0.006 | $F_{1,1248.6}$=0.101 | 0.750 |
| | $R^2_{cond}$: 0.385; $R^2_{mar}$: 0.138 | | | | $R^2_{cond}$: 0.472; $R^2_{mar}$: 0.062 | | | |
| **Yellow breast patch brightness (n. obs.: Corsica=3890; D-Rouvière=2007)** | | | | | | | | |
| **Fixed effects** | Est | SE | F | P | Est | SE | F | P |
| Year | **-1.294** | **0.310** | **$F_{1,13.11}$=21.860** | **<0.001** | **-1.481** | **0.480** | **$F_{1,13.09}$=9.706** | **0.008** |
| Sex(fem) | **0.627** | **0.090** | **$F_{1,1904.7}$=48.007** | **<0.001** | 0.503 | 0.127 | **$F_{1,935.33}$=15.581** | **<0.001** |
| Year*sex(fem) | **-0.286** | **0.089** | **$F_{1,2600.6}$=10.271** | **0.001** | -0.004 | 0.126 | $F_{1,1169.5}$=0.001 | 0.971 |
| | $R^2_{cond}$: 0.391; $R^2_{mar}$: 0.199 | | | | $R^2_{cond}$:0.507; $R^2_{mar}$: 0.169 | | | |

## Climate across years

In Corsica we found a significant increase in average temperature (1.23ºC; 0.080±0.024, $F_{1,13}$=10.782, p=0.005) and decrease in average precipitation (-0.64mm; -0.068±0.021, $F_{1,13}$=10.306, p=0.006) during moult across the last 15 years (Fig. 2). In D-Rouvière, while the two climatic variables showed similar trajectories than in Corsica, there was no significant change over time neither in average temperature (0.049±0.037, $F_{1,13}$=1.777, p=0.205) nor in precipitation (-0.043 ±0.056, $F_{1,13}$=0.601, p=0.452) during moult (Fig. 2).

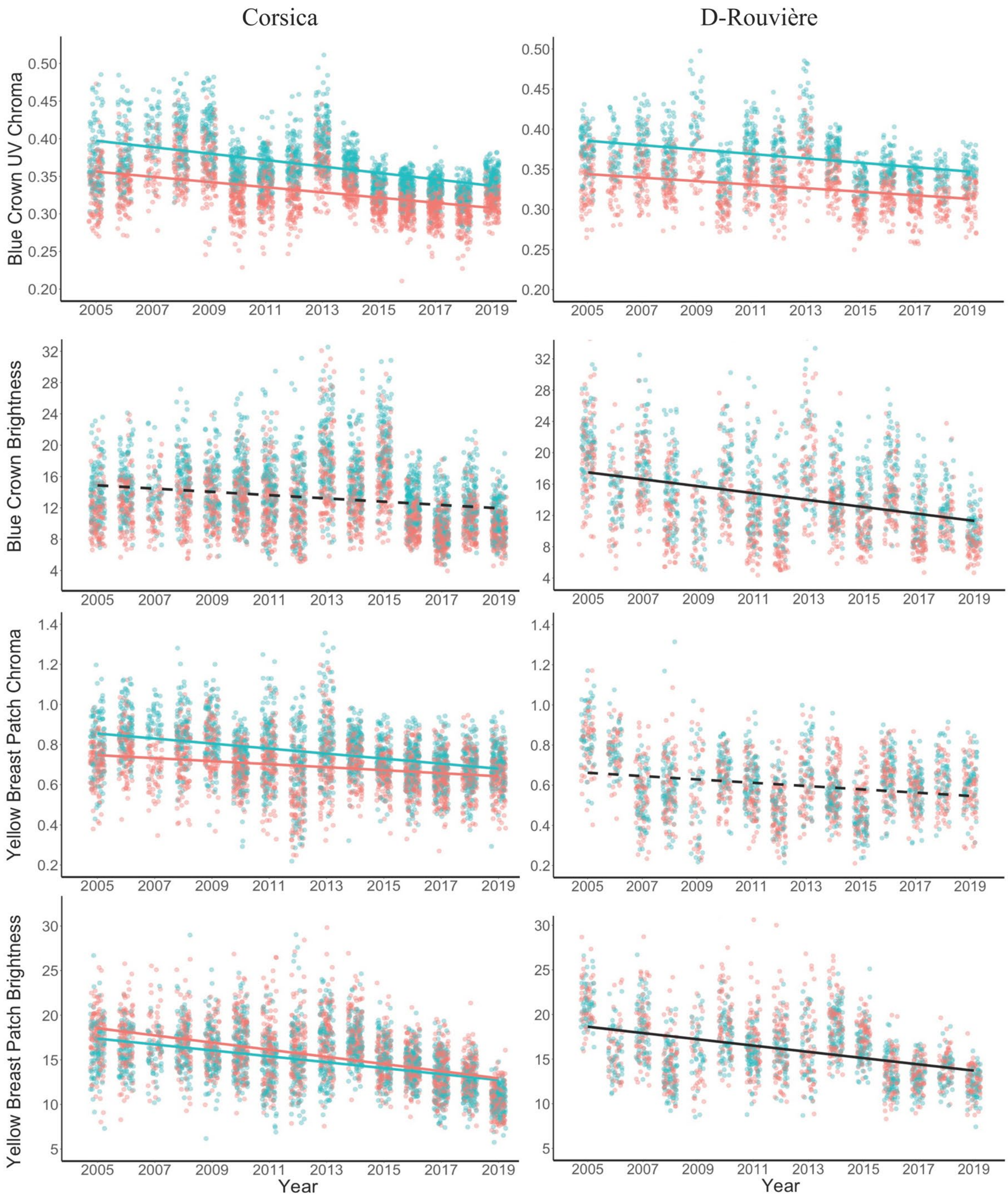

Figure 1: Linear decrease in male and female blue crown and yellow breast patch colouration in blue tits of Corsica and D-Rouvière across time. Blue dots represent males and red dots females. The lines show the predicted slope values (Table 1). Black lines show the directional changes in colour across time when the associations were not sex dependent. Blue and red lines show the directional change in colour across time for males and females respectively when the trajectories were sex dependent. Dashed lines represent non-significant associations. A jitter was added to visualize overlapping data points.

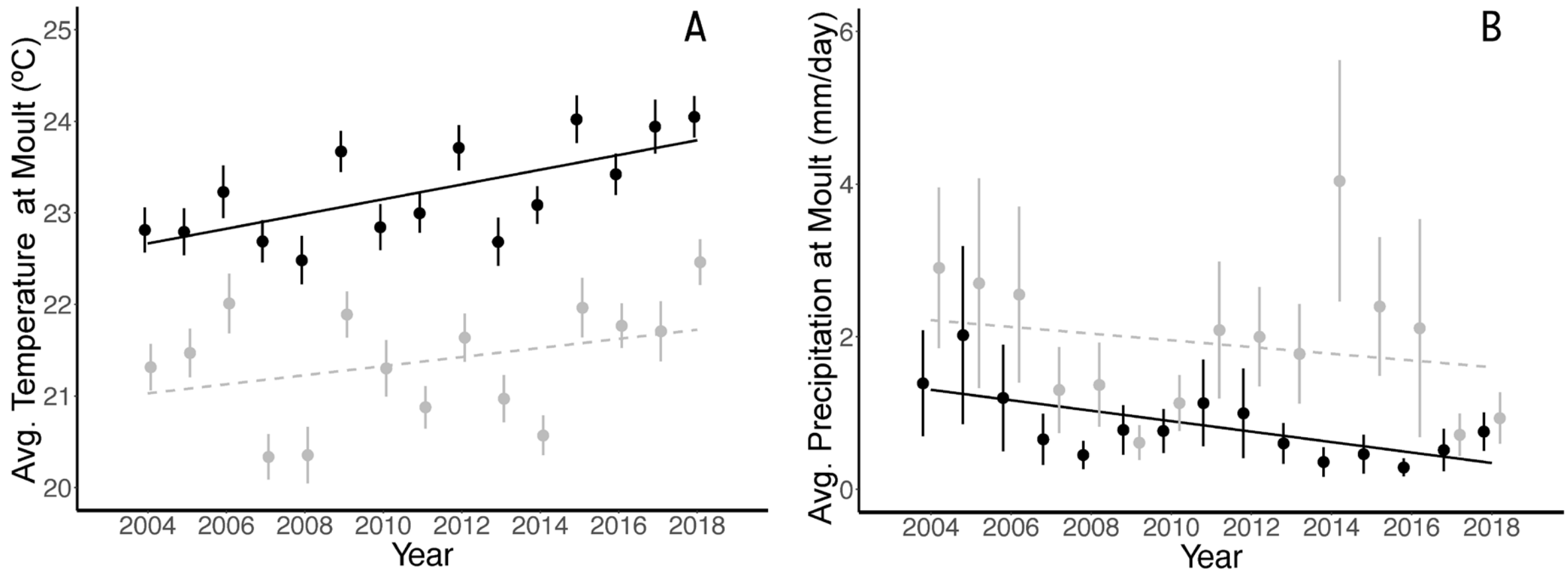

Figure 2: Change in average temperature (A) and precipitation (B) during moulting along the study period. Black dots and black solid lines represent the Corsican population and grey dots and grey dashed lines the D-Rouvière population. Only the Corsican associations are significant.

**Association between the coloured traits and the climatic variables**

**Blue crown**: In Corsica, for the UV chroma we found weak but significant interactions between sex and both average temperature and precipitation during moulting suggesting sex-dependent negative associations with temperature and positive with precipitations ($R^2_{mar}$: 0.308; $R^2_{cond}$: 0.750, Table 2). When analysing each sex separately, the negative associations between UV chroma and average temperature during moult were only marginally significant in males (-0.012±0.006, $F_{1,11.93}$=3.781, p=0.075, Fig. 3) and females (-0.010±0.005, $F_{1,11.93}$=3.180, p=0.099, Fig. 3). Within each sex separately, no significant associations between average precipitations during moult and UV chroma in males (0.007±0.006, $F_{1,11.97}$=1.715, p=0.214) or females (0.005±0.005, $F_{1,11.98}$=0.930, p=0.353) appeared. In D-Rouvière, we found no significant associations between the UV chroma and the climatic variables (Table 2). For brightness, our results also showed no significant associations with the average temperature in Corsica but a marginal positive association in D-Rouvière (see Fig. S9.1). No significant associations were found between brightness and average precipitations during moult (Table 2).

**Yellow breast patch:** In Corsica, we found that yellow chroma was negatively associated with average temperature and positively associated with precipitation during moult (Table 2), and that these associations differed between males and females ($R^2_{mar}$: 0.133; $R^2_{cond}$: 0.395, Table 2, Fig. 3). When we analysed the interactions in each sex separately, in males we found that yellow chroma decreased with higher temperatures (-0.038±0.014, $F_{1,11.82}$=6.827, p=0.022), and that it increased when precipitations were more abundant (0.025±0.013, $F_{1,12.04}$=3.452, p=0.087), although the latter relationship was marginally significant. In females, yellow chroma also decreased with higher temperatures but with a shallower slope than in males (-0.025±0.013, $F_{1,11.96}$=3.416, p=0.089). In females, no significant relationship was found between chroma and average precipitation during moulting (0.009±0.013, $F_{1,12.14}$=0.516, p=0.486, Fig. 3). In D-Rouvière, a sex-dependent association between chroma and both average temperature and precipitation during moulting was also found. Sex-specific analyses revealed a marginal positive association between average temperature at moult in males (0.054±0.024, $F_{1,10.98}$=4.749, p=0.052) and females (0.037±0.020, $F_{1,11.04}$=3.436, p=0.090, see SM-8 Fig. S1). No significant associations were found for the association between chroma and average precipitations in males (-0.011±0.024, $F_{1,10.98}$=0.232, p=0.639) or females (0.006±0.019 $F_{1,10.97}$=0.011 p=0.745). Last, in Corsica, for brightness our results showed a negative association with average temperature during moulting in both males and females and a significant interaction between average precipitation during moulting and sex ($R^2_{mar}$=0.163; $R^2_{cond}$:0.412, Table 2). After analysing each sex separately, we found no significant associations between yellow brightness and precipitations during moulting in males (0.553±0.442, $F_{1,13.10}$=1.564, p=0.232) or females (0.873±0.468, $F_{1,13.083}$=3.477, p=0.084). No significant associations were found for yellow brightness in D-Rouvière (see Tables S8.2 and S8.4 for further details).

**Table 2**. Associations between the four colour components and the average temperature (Avg. temp.) and average precipitation (Avg. prec.) during moulting in Corsica and D-Rouvière. Significant (p<0.05) variables are in bold. $R^2_{cond}$ represents the variance explained by both the fixed and random effects included in the model, $R^2_{mar}$ represents the variance explained by the fixed factors alone.

| | **Corsica** | | | | **D-Rouvière** | | | |
|---|---|---|---|---|---|---|---|---|
| **Blue crown UV chroma (n. obs.: Corsica=3867; D-Rouvière=2068)** | | | | | | | | |
| **Fixed Effects** | Est | SE | F | P | Est | SE | F | P |
| Avg. temp. | -0.012 | 0.006 | $F_{1,12.0}$=3.537 | 0.084 | -0.003 | 0.005 | $F_{1,10.98}$=0.306 | 0.590 |
| Avg. prec. | 0.007 | 0.005 | $F_{1,12.0}$=1.343 | 0.268 | -0.003 | 0.005 | $F_{1,10.98}$=0.493 | 0.496 |
| Sex(fem) | **-0.035** | **0.0008** | **$F_{1,2205.4}$=1609.5** | **<0.001** | **-0.038** | **0.001** | **$F_{1,1185.9}$=1030.5** | **<0.001** |
| Year | | | | | -0.010 | 0.005 | $F_{1,11.02}$=3.688 | 0.081 |
| Avg. temp.*sex(fem) | **0.001** | **0.0007** | **$F_{1,3704.5}$=4.927** | **0.026** | 0.001 | 0.001 | $F_{1,1986.0}$=1.696 | 0.192 |
| Avg. prec.*sex(fem) | **-0.002** | **0.0008** | **$F_{1,3838.0}$=10.46** | **0.001** | 0.0005 | 0.001 | $F_{1,2039.3}$=0.261 | 0.609 |
| | $R^2_{cond}$: 0.750; $R^2_{mar}$: 0.308 | | | | $R^2_{cond}$: 0.709; $R^2_{mar}$: 0.338 | | | |
| **Blue crown brightness (n. obs.: Corsica=3867; D-Rouvière=2068)** | | | | | | | | |
| **Fixed Effects** | Est | SE | F | P | Est | SE | F | P |
| Avg. temp. | -0.765 | 0.661 | $F_{1,12.0}$=1.010 | 0.334 | 1.414 | 0.670 | $F_{1,10.97}$=4.811 | 0.051 |
| Avg. prec. | 0.039 | 0.615 | $F_{1,12.0}$=0.008 | 0.927 | 0.763 | 0.661 | $F_{1,10.95}$=1.260 | 0.285 |
| Sex(fem) | **-2.379** | **0.118** | $F_{1,2118.3}$=400.85 | <0.001 | **-2.325** | **0.189** | **$F_{1,1108.0}$=149.87** | **<0.001** |
| Year | | | | | **-2.015** | **0.666** | **$F_{1,10.99}$=9.144** | **0.011** |
| Avg. temp.*sex(fem) | 0.206 | 0.117 | $F_{1,3833.5}$=3.109 | 0.077 | 0.084 | 0.181 | $F_{1,2036.1}$=0.218 | 0.640 |
| Avg. prec.*sex(fem) | 0.035 | 0.118 | $F_{1,3789.7}$=0.089 | 0.765 | -0.057 | 0.183 | $F_{1,2045.9}$=0.098 | 0.753 |
| | $R^2_{cond}$:0.444; $R^2_{mar}$: 0.094 | | | | $R^2_{cond}$: 0.511; $R^2_{mar}$: 0.218 | | | |
| **Yellow breast patch chroma (n. obs.: Corsica=3890; D-Rouvière=2007)** | | | | | | | | |
| **Fixed Effects** | Est | SE | F | P | Est | SE | F | P |
| Avg. temp. | **-0.038** | **0.013** | **$F_{1,12.0}$=5.532** | **0.036** | 0.052 | 0.022 | $F_{1,11.01}$=4.272 | 0.063 |
| Avg. prec. | 0.025 | 0.012 | $F_{1,12.1}$=1.918 | 0.191 | -0.011 | 0.021 | $F_{1,10.99}$=0.009 | 0.922 |
| Sex(fem) | **-0.068** | **0.004** | **$F_{1,2070.1}$=276.90** | **<0.001** | **-0.012** | **0.006** | **$F_{1,1022.9}$=3.979** | **0.046** |
| Year | | | | | **-0.055** | **0.021** | **$F_{1,11.07}$=6.725** | **0.024** |
| Avg. temp.*sex(fem) | **0.013** | **0.004** | **$F_{1,3850.4}$=10.590** | **0.001** | **-0.012** | **0.005** | **$F_{1,1964.2}$=4.398** | **0.036** |
| Avg. prec. *sex(fem) | **-0.015** | **0.004** | **$F_{1,3816.5}$=13.514** | **<0.001** | **0.018** | **0.005** | **$F_{1,1989.0}$=9.713** | **0.001** |
| | $R^2_{cond}$: 0.395; $R^2_{mar}$=0.133 | | | | $R^2_{cond}$: 0.492; $R^2_{mar}$=0.137 | | | |
| **Yellow breast patch brightness (n. obs.: Corsica=3890; D-Rouvière=2007)** | | | | | | | | |
| **Fixed Effects** | Est | SE | F | P | Est | SE | F | P |
| Avg. temp. | **-1.050** | **0.408** | **$F_{1,12.0}$=7.654** | **0.017** | 0.154 | 0.562 | $F_{1,11.03}$=0.055 | 0.817 |
| Avg. prec. | 0.310 | 0.380 | $F_{1, 12.1}$=1.389 | 0.261 | 0.577 | 0.545 | $F_{1,11.01}$=0.874 | 0.369 |
| Sex(fem) | **0.625** | **0.090** | **$F_{1,1898.2}$=47.904** | **<0.001** | **0.501** | **0.127** | **$F_{1,941.90}$=15.501** | **<0.001** |
| Year | | | | | **-1.426** | **0.536** | **$F_{1,11.05}$=7.062** | **0.022** |
| Avg. temp.*sex(fem) | -0.143 | 0.089 | $F_{1,3858.7}$=2.560 | 0.109 | -0.044 | 0.124 | $F_{1,1982.5}$=0.128 | 0.719 |
| Avg. prec.*sex(fem) | **0.269** | **0.090** | **$F_{1,3797.1}$=8.899** | **0.002** | -0.142 | 0.125 | $F_{1,1975.4}$=1.290 | 0.256 |
| | $R^2_{cond}$: 0.412; $R^2_{mar}$=0.163 | | | | $R^2_{cond}$:0.524; $R^2_{mar}$=0.172 | | | |

## Quantitative genetic analyses

We found low to moderate heritabilities for the coloured traits in our studied populations, ranging from 0.026 to 0.167 in Corsica and from 0.029 to 0.173 in D-Rouvière (Table 3). Consistent environmental differences across individuals (permanent environment) accounted for a small fraction of the total phenotypic variance (Corsica range: 0.6% – 5.1%, average: 2.6% and D-Rouvière: range: 1.6% – 6.6%, average: 4.2%) that was mostly explained by year (Corsica range: 15.0% - 43.8%, average: 28.1% and D-Rouvière: range 28.5% - 46.0%, average: 37.0%; for more information on each random factor variance see Tables S10.1-S10.6). Regarding the temporal trends in breeding values, most of the posterior mode estimates were negative, yet in all of them the 95% credible interval (CI) included zero (Table 4) hence there was no evidence for a temporal trend in breeding values for any of the traits/areas (see Fig. S11.1 and S11.2).

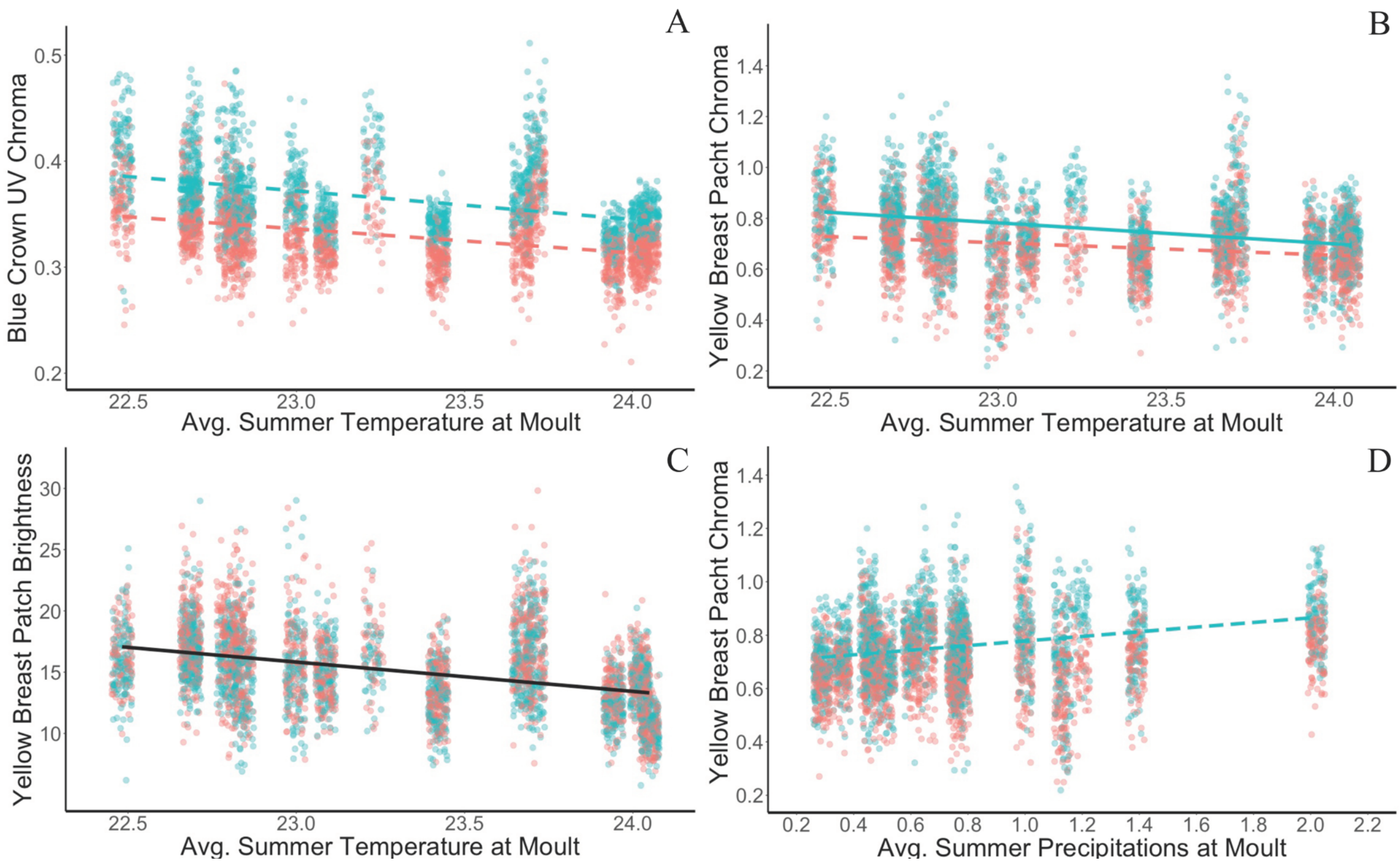

Figure 3: In Corsica, blue crown UV chroma (A), yellow breast patch brightness (B) and chroma (C) decreased with increased average temperatures during moulting (see Table 2). In addition, yellow breast patch chroma (D) increased with higher average precipitations during moulting. Blue dots and lines represent males and red ones females. The black lines represent the associations when there was no interaction with sex. Solid lines represent significant results, dashed lines marginally significant relationships. A jitter was added to visualize overlapping data points.

**Table 3**: Heritability ($V_A/V_P$) and 95% Credible Intervals (CI) obtained from the animal models of each coloured trait in each sex and population.

| Corsica | | | | |
|---|---|---|---|---|
| | Males | | Females | |
| | $h^2$ | 95% CI | $h^2$ | 95% CI |
| Blue crown UV chroma | 0.167 | [0.088, 0.237] | 0.161 | [0.086, 0.233] |
| Blue crown brightness | 0.029 | [<0.001, 0.070] | 0.053 | [<0.001, 0.103] |
| Yellow breast patch chroma | 0.074 | [<0.001, 0.143] | 0.026 | [<0.001, 0.074] |
| Yellow breast patch brightness | 0.078 | [<0.001, 0.134] | 0.034 | [<0.001, 0.079] |
| **D-Rouvière** | | | | |
| | Males | | Females | |
| Blue crown UV chroma | 0.058 | [<0.001, 0.128] | 0.173 | [0.063, 0.266] |
| Blue crown brightness | 0.068 | [<0.001, 0.135] | 0.073 | [<0.001, 0.132] |
| Yellow breast patch chroma | 0.110 | [0.016, 0.193] | 0.080 | [<0.001, 0.151] |
| Yellow breast patch brightness | 0.039 | [<0.001, 0.097] | 0.029 | [<0.001, 0.088] |

**Table 4:** Posterior mode estimates and CI (95%) of the linear regression of the BLUPs obtained from the animal models against the mean year.

| Corsica | | | | |
|---|---|---|---|---|
| | Males | | Females | |
| | Estimate | 95% CI | Estimate | 95% CI |
| Blue crown UV chroma | -0.0014 | [-0.0043, 0.0012] | -0.0002 | [-0.0036, 0.0010] |
| Blue crown brightness | 0.0107 | [-0.2155, 0.1548] | -0.0012 | [-0.2429, 0.1255] |
| Yellow breast patch chroma | -0.00005 | [-0.0100, 0.0060] | -0.00002 | [-0.0043, 0.0044] |
| Yellow breast patch brightness | 0.0043 | [-0.1738; 0.1791] | -0.0004 | [-0.1306, 0.099] |
| **D-Rouvière** | | | | |
| | Males | | Females | |
| Blue crown UV chroma | -0.00003 | [-0.0003, 0.0023] | -0.0010 | [-0.0055; 0.0019] |
| Blue crown brightness | -0.00009 | [-0.6115, 0.3825] | 0.0332 | [-0.5408; 0.2943] |
| Yellow breast patch chroma | -0.0011 | [-0.0208, 0.0180] | -0.0007 | [-0.0156; 0.0125] |
| Yellow breast patch brightness | -0.0012 | [-0.1591, 0.3134] | 0.0009 | [-0.1935; 0.2288] |

## Discussion

Our longitudinal 15-year data set including more than 5800 colour ornament measures shows that the yellow breast patch and blue crown colourations of male and female Mediterranean blue tits have become duller and less chromatic throughout the years (decrease from 9% to 23% see Table S12.1). Over the same period, we detected an increase in temperature and a decrease in precipitation, mainly in Corsica. Associating the climatic variation with colouration shows that colour changes were partly associated with changes in temperature and precipitation during moulting particularly in the more southern Corsican population. Lastly, we found that the temporal trends described at the phenotypic level were not aligned with a change at the genetic level, evidencing that the described change did not represent a microevolutionary response but rather originated from plasticity. Overall, these results suggest an impact of climate change on the visual communication system of the blue tits. Given the potential role of these traits in social and sexual selection and the importance of mate choice for adaptation (Whitlock and Agrawal 2009; Gómez-Llano et al. 2021) our results raise the question of whether this reduction could affect species abilities to react to climate change.

**Temporal trends and association with the climatic proxies**

The studies exploring the association between colouration and climatic variables have traditionally focused on melanin-based traits in several taxa (e.g. Gloger rule or thermal melanism hypothesis; Delhey 2017), including birds (Fargallo et al. 2018; Delhey et al. 2019), reptiles (Martínez-Freiría et al. 2020) or insects (Zeuss et al. 2014; Clusella-Trullas and Nielsen 2020). Still, most of this work has not explored the impact of climate change on ornamental colourations, despite their importance for fitness due to their role as signals in different intra- and inter-sexual contexts (Kodric-Brown 1985; Stuart-Fox and Ord 2004; Girard et al. 2015; López-Idiáquez et al. 2016*b*). The evidence available, however, has shown that the expression of achromatic and melanin-based ornaments can be driven by climate change (Evans and Gustafsson 2017; Moore et al. 2021). Here, our results suggest that rapid climate change can in addition impact both structural and carotenoid-based colourations (the other two major colouration types) in the blue tits from Corsica. However, it should be noted that collinearity exists between year and climate, and caution is needed: other environmental factors might cause this association as well, calling for further studies to confirm it.

In our study, warmer and dryer summers in Corsica led to a reduced chromaticity of the UV blue crown colouration. This result is in line with a previous study in the burrowing parrot (*Cyanoliseus patagonus*) showing that feathers grown in dryer years displayed less bright structural colours (Masello et al. 2008). While unravelling the physiological processes behind these associations will require specific studies, a potential driver could be a reduction in food availability or quality during hot and dry summers (Both 2010). An effect that would fit with the results of a meta-analysis including information on birds, insects, reptiles and arachnids and supporting the condition-dependence of structural colourations (White 2020). Further, brood size manipulation experiments in wild blue tits and eastern bluebirds (*Sialia sialis*) also seem to support this idea suggesting that food limitations can constrain the expression of the offspring's blue UV colourations (Jacot and Kempenaers 2007; Siefferman and Hill 2007), although this association was not always found in captivity or in adults (McGraw et al. 2002; Peters et al. 2011). Besides, a non-exclusive alternative could be that hot and dry summers lead to an increased occurrence of energetically costly behaviours that allow heat dissipation (du Plessis et al. 2012; Pattinson et al. 2020; McKechnie et al. 2021). While experimental studies are needed to clarify the mechanism behind the association of climate and structural colouration, our results seem to support the suspected (Jacot and Kempenaers 2007; Siefferman and Hill 2007; Doutrelant et al. 2012), but debated (Prum 2006) and not often demonstrated environmental sensitivity of structural colourations (Masello et al. 2008).

Regarding the influence of climate on carotenoid-based traits the information available is more abundant but is also mixed. A recent comparative analysis has shown that in nine Australian bird species the individuals living in warmer areas had more saturated carotenoid colourations than those living in colder ones, while the opposite pattern (less saturation in warmer areas) was found in three species (Prasetya et al. 2020). Besides, in guppies, an experimental study showed a non-linear association between their orange colouration hue and temperature, with enhanced colourations at intermediate temperatures (Breckels and Neff 2013). Finally, two long-term studies have explored the associations between climate and the carotenoid-based colourations showing that, while higher temperature and lower precipitation at moult led to a reduction in the red tail colouration of the American redstart (*Setophaga ruticilla)* in Canada (Reudink et al. 2015) they, by contrast, enhanced the yellow chest colouration in Hungarian great tits (Laczi et al. 2020). In both studies the reported patterns were ascribed to variations in food availability, suggesting a potential opposite effect of climate on food availability or quality depending on the study area. Here we found that temperature was negatively correlated with yellow breast patch chroma and brightness and that there was a marginal positive association between precipitations and yellow breast patch chroma in males. Therefore, our results for the yellow breast patch colouration strongly differ from those found by Laczi et al. (2020) in great tits. Mediterranean areas undergo hotter and drier summers compared to Hungary so we hypothesize that this lack of consistency may be because Mediterranean birds face a reduction in food availability during moult or other heat-related costs (du Plessis et al. 2012; Gardner et al. 2016; Pattinson et al. 2020; McKechnie et al. 2021) not faced by the Hungarian birds. If true, the impacts of warming on ornamental traits could be more severe in warmer areas within a species distribution. Further comparison across more species will in the future allow determining whether the documented variation between species and study areas is a common phenomenon.

Finally, our results agree with those previously published reporting negative effects of climate change on ornamental traits (Svensson 2019; Moore et al. 2021, but see Møller and Szép 2005). It is interesting to highlight, however, that most of the published evidence comes from a small number of species, mostly birds (Svensson 2019). Considering that the consequences of climate change may differ depending on the species and their particular constraints (e.g. ectotherms may be differently affected than endotherms; Aragón et al. 2010), studies encompassing different taxa are needed to grab a general understanding of the consequences of climate change on ornamental traits.

**Quantitative genetics of the blue tit colourations**

Based on a social pedigree across 16 generations including all birds with colour measures, quantitative genetic models report that the blue and yellow colouration of the blue tits present low to moderate heritabilities, concordant with estimates previously published in this species. Repeatabilities were also low, suggesting that the studied traits are highly dynamic. The low repeatability, along with the fact that year accounted for much of the phenotypic variance, points out that environmental conditions play an important role as drivers of the expression of the blue and yellow colours in the blue tit, in agreement with our results suggesting that climate influences their colouration.

Despite low to moderate heritabilities, the existence of additive genetic variance on the blue tit colourations raised the possibility that the trends described at the phenotypic level were caused by a change at the genetic level. However, no significant temporal trends in the breeding values were detected for any of the studied traits, neither in males or females in Corsica or D-Rouvière. While such analyses on breeding value trends are fundamental to test for evidence of microevolution, they are scarcely conducted because of the lack of appropriate data and the complexity to account for uncertainty around breeding value estimation (Hadfield et al. 2010). For this reason, it still remains virtually unknown whether climate change is leading to a change at the genetic level in ornamental traits. So far, only one study has explored this issue showing a reduction in the breeding values for the forehead patch of the collared flycatcher (Evans and Gustafsson 2017). Interestingly, our results contrast with these as we did not find a change at the genetic level. This difference may be explained by higher heritability for the sizes of melanin/white colour patches, ranging between 0.35 and 0.79 (in species such as pied and collared flycatchers, garter snakes (*Thamnophis sirtalis*) or guppies; Qvarnström 1999; Brooks and Endler 2001; Hegyi et al. 2002; Westphal and Morgan 2010; Potti and Canal 2011; Evans and Gustafsson 2017), compared to heritabilities of structural or carotenoid-based traits, ranging from 0.02 to 0.25 (in blue and great tits and guppies, for example; Brooks and Endler 2001; Hadfield et al. 2006; Evans and Sheldon 2012; Drobniak et al. 2013; Charmantier et al. 2017). Because of this low heritability, the maximum expected evolutionary response could be small, and we may lack power to detect it. In any case, the non-significance of the temporal trends in breeding values strongly suggests that the documented phenotypic decline over time is caused by a plastic response to the environmental conditions.

**Differences in the climatic effects for *ogliastrae* and *caeruleus* subspecies**

Previous studies have reported that the strength and direction of the responses to climate change can be population-dependent with, for example, stronger phenological responses at higher latitudes (Parmesan 2007) or near the edge of a species range (Sheth and Angert 2016). Further, climate dependent selection can lead to population differences in colouration. For instance, in Northern and Central Italy, wall lizards (*Podarcis muralis*) display more conspicuous green and black colorations in hotter and drier environments (Miñano et al. 2021). Here we found that while three of the four studied colour components (blue crown UV chroma and yellow breast patch brightness and chroma) were negatively associated with temperature during moult in Corsica, in D-Rouvière two of them (blue crown brightness and yellow breast patch chroma) showed marginal positive associations with temperature, suggesting a differential response to warming in the two populations. This difference could be explained by the fact that the effect of climate change on temperature and rain was more marked in Corsica (*ogliastrae* subspecies) than in D-Rouvière (*caeruleus* subspecies). In line with this, a study in the same populations showed that warming in spring was associated with the timing of laying of the blue tits in Corsica but not in D-Rouvière (Bonamour et al. 2019). This difference could also be explained because climate in Corsica is hotter and dryer when compared to D-Rouvière, and thus it is likely that Corsican blue tits are closer to their thermal limits. If the latter is true, the raise in temperature could have different effects in Corsica and D-Rouvière, given that latitude is not expected to have a large impact on the heat resistance of endotherms (Araújo et al. 2013). For instance, in *Drosophila* the association between heat resistance and latitude (Kimura 2004) or temperature across their distribution range is weak (Kellermann et al. 2012). From a more mechanistic perspective, the differences in the temperature dependence among our two populations could be caused by higher humidity in D-Rouvière than in Corsica. Studies in fruit flies have shown that the effects of temperature may be mitigated in more humid regions (i.e. where the loss of water may be less important; Kellermann et al. 2012). Whether this is happening in birds needs to be confirmed, but it highlights the important role precipitations may play as drivers of selection (Siepielski et al. 2017) and as modulators of the effects of other environmental variables on ornamental and other types of traits.

The lack of association between blue tit colours and climate in D-Rouvière leaves however one question open: why is there a decrease in colouration in this population? Several hypotheses can be formulated. For instance, blue tit colour changes over time may not be directly generated by climate but by other factors that also change with time and that have an impact on colouration, such as habitat structure (Medina et al. 2017), parasite prevalence (Janas et al. 2018) or food availability that may have decreased due to the increase in the use of pesticides (Møller et al. 2021). Alternatively, it is also possible that our climate windows and variables were more efficient at capturing the variability in temperature and precipitation in Corsica than in D-Rouvière where other variables and windows may be more relevant. However, although we could not point the exact environmental variable, the described changes in colour are nonetheless plastic as we could not detect a change in the breeding values.

**Sex differences in the temporal and climatic trends**

We found sex-dependent associations in both populations, with overall stronger trends in males than in females, suggesting a higher environmental sensitivity of male colouration. This finding opens exciting perspectives to explore the mechanisms explaining this sex-specific pattern. Male blue tits face stronger inter-sexual selection than female blue tits (Doutrelant et al. 2020), and thus, it is possible that they invest more in ornamentation than females. This could explain why it has been described in Swedish blue tits that males start moulting earlier than females, even when there are still young on the nest (Svensson and Nilsen 1997). As a consequence of this higher investment in ornamentation, warming or changes in precipitation regimes could have a stronger impact on male colouration via their influences on male body condition or on parasite prevalence. Alternatively, because of sex-specific investment and sex-specific resulting costs (Fitzpatrick et al. 1995), female condition may be more sensitive to the costs inherent to reproduction, such as egg production (Doutrelant et al. 2012) than to those imposed by climate at moult, and thus the climatic effects occurring at moulting time may be milder in this sex.

**Potential consequences of the documented reduction in colouration**

Whether sexual selection facilitates or hinders the adaptation to novel environmental conditions is still debated (Candolin and Heuschele 2008). On the one hand, sexual selection could promote adaptation, for instance, by favouring higher “quality” individuals (Hamilton and Zuk 1982; Whitlock and Agrawal 2009; Gómez-Llano et al. 2021). On the other hand, sexual selection might have detrimental demographic consequences, for example, by favouring the presence of costly traits whose expression negatively influences the viability of the individuals (Long et al. 2009). Here, our data suggests a reduction in the colouration of the blue tits over time (mean reduction of 15.51%, see Table S12.1), and a reduction in the variance of some of the studied traits (e.g., blue UV and yellow chroma in Corsica and blue UV chroma in D-Rouvière, see SM-13). Given the potential role of the blue tit colouration as a secondary sexual trait, our results may be evidencing a decrease in the signalling potential of its colour ornaments. For example, if there is not enough variance among males, females will not be able to discern between high- and low-quality individuals. Such effect has been found in the superb fairy-wren (*Malurus cyaneus*), whereby males cannot acquire their breeding plumage in time when dry summers occur, resulting in weaker sexual selection (Cockburn et al. 2008). The importance of such loss of signalling potential is highlighted by a recent experiment in *Drosophila melanogaster* concluding that mating regimes with strong sexual selection led to increased female fitness when compared to regimes lacking sexual selection under a warming scenario (Gómez-Llano et al. 2021). If blue tits are experiencing a decreasing strength in sexual selection for colour ornaments, this could hence result in a reduced capacity for these populations to adapt to the new conditions imposed by climate change. Further studies are needed to explore this more deeply, ideally including other signalling traits present in this species, as if females cannot rely on colour during mate choice selection could favour alternative signalling systems, like male song.

## Conclusions

Our data suggests that in the last 15 years there has been a general decrease in the colouration of blue tits breeding in two Mediterranean populations. In Corsica, where climate has become warmer and drier across the study period, the changes in colouration were related to changes in temperature at moult, suggesting a connection with climate change. Also, our results discard the presence of a microevolutionary change in blue tit colouration over the past 15 years. Based on this evidence, future research should aim at analysing the link between climate and ornaments in other bird species and taxa, elucidating the mechanisms linking climate to the observed phenotypic change and the consequences of the reported trends for sexual selection and adaptation to climate change.

**Acknowledgements**

We thank all the researchers involved in the long-term monitoring of the blue tit populations, and in particular: Denis Réale, Philippe Perret, Jacques Blondel, Samuel Caro, and Marcel Lambrechts, and Annick Lucas, Pablo Giovannini, Samuel Perret from the PLT. We thank Doris Gomez for all her expertise on colour measurement. We recognize the help of all the students and field assistant involved in capturing the blue tits throughout the years and those that participated measuring the coloration, in particular Emeline Mourocq and Afiwa Midamegbe. We thank the ONF, the APEEM and the mayors of Manso and Montarnaud for allowing us to work on E-Pirio and D-Rouvière and the landowners of D-Muro and E-Muro for their permission to work on their properties. The comments from J. Lau, E. Svensson and three anonymous reviewers greatly improved the previous versions of this work. The long-term monitoring was supported by the OSU-OREME, several ANR to A. Charmantier, C. Doutrelant and C. Teplitsky (in particular ANR 09-JCJC- 0050, ANR-12-ADAP-0006-02-PEPS), the Languedoc-Rousillon region '*chercheur d'avenir*' grant to C. Doutrelant and the ERC 2013-StG-337365-SHE to A. Charmantier. D. López-Idiáquez was funded by a postdoctoral grant of the Basque Country Government (POS_2019_1_0026).

**Data and Code Accessibility:**

Data and code underlying this article have been deposited in the Dryad Digital Repository (https://doi.org/10.5061/dryad.w6m905qr9; López-Idiáquez et al. 2022)

***Supplementary material***

# *Long-term decrease in colouration: a consequence of climate change?*

## SM-1: Feather sampling, colouration measurement and sample sizes

Feather sampling took place following a standardised protocol that remained unchanged during the all the study period. Specifically, six blue feathers were taken from the front central part of the head crown (right side during odd years and left side in even years) and 8 yellow feathers from the upper part of the chest (right side during odd years and left side in even years). Feather sampling was conducted mostly by thirteen permanent staff (54% of the samples [3207/5977]) and PhD students (34% of the samples [2050/5977]) that are highly trained and experienced in bird sampling and in the way the feathers need to be plucked. Thus, from the total of 5977 individuals sampled 5257 (88%) were sampled by highly trained researchers that have been involved in the fieldwork several of the study years.

To measure the colouration, we used a spectrophotometer (AVASPEC-2048, Avantes BV, Apeldoorn, Netherlands) and a deuterium-halogen light source (AVALIGHT-DH-S lamp, Avantes BV) covering the range 300-700 nm (Doutrelant et al. 2008; 2012) and kept at a constant angle of 90º from the feathers. Specifically, we used two different lamps, same brand, model, and characteristics (lamp 1 from 2005 to 2015 and from 2018 to 2019, and lamp 2 from 2016 to 2017) in our measurements. For each bird and colour patch, we computed the mean of six reflectance spectra taken on two sets of three blue and four yellow feathers (Doutrelant et al. 2008; 2012; see SM-1 Table 1 for further detail on the sample sizes). Following previous studies (Doutrelant et al. 2008; 2012; Andersson et al. 1998), we computed chromatic and achromatic colour variables for our study colourations based on the shape of the spectra using Avicol v2 or the R package "*pavo*" (Gomez 2006; Maia et al. 2019). For the blue crown coloration, we computed one chromatic variable: UV chroma (proportion of the total reflectance falling in the range 300-400 nm) and one achromatic variable: brightness (area under the reflectance curve divided by the width of the interval 300-700 nm). For the yellow breast patch colouration, we computed chroma as $(R_{700}-R_{450})/R_{700}$, with higher values of carotenoid chroma being linked to higher carotenoid contents in the plumage (Isaksson et al. 2008), in addition to brightness.

All the measurements were taken by people trained by either Claire Doutrelant or Doris Gómez, who have ample experience in taking colouration measurements using spectrophotometers. Most measurements were taken by a single person, María del Rey with some exceptions. In D-Rouviere, María del Rey measured the feathers sampled in ten years, the remaining five being measured by A. Rieux (2005), N. Grnac (2006 and 2007), A. Midamegbe (2008), and A. Fargevielle (2015). The measurements of the feathers were conducted the same year the samples were taken, with the exception of those sampled in 2009 and 2010 that were measured in 2014, and those sampled in 2016 and 2017 that were measured in 2017 and 2018 respectively. In Corsica, Maria del Rey measured all the feathers sampled in 8 years (2010, 2012-2015, and 2017-2019), in 6 years (2005-2009 and 2011) María del Rey and E. Mourocq measured 1/3 of the feathers each, being the other measurers A. Fargevieille (15 inds. per year), L. Chassagne (2005), S. Robert (2006) and M. Thion (2011). Finally, the year 2016 was fully measured by A. Fargevieille. Here, as in D-Rouviere, measures were taken the same year the feathers were sampled, with the exception of four years. The samples of 2010 and 2013 were measured in 2014, and the samples of 2016 and 2017 were measured the years 2017 and 2018 respectively.

The extraction of the colour variables to be used in the analyses from the colour spectra was done by a single researcher, A. Fargevieille, following the same approach in all years. Also, before extracting the colour variables from the mean spectra the repeatability of the colour variables was checked, both within- and among-individuals. Whenever a low repeatability at the among-individual level was detected the spectra of the individuals with higher within-individual variability was checked. If one or two spectra out of the six taken were different from the others they were removed. If there were more than two spectra looking different or they were not consistent the feathers of those individuals were remeasured. This cleaning process was repeated until no aberrant spectra were found.

Table S1.1: Number of observations per coloured trait, sex and year in Corsica and in D-Rouvière (mainland).

| | Blue crown colouration | | | | Yellow breast patch colouration | | | |
|---|---|---|---|---|---|---|---|---|
| | Corsica | | D-Rouvière | | Corsica | | D-Rouvière | |
| Year | Males | Females | Males | Females | Males | Females | Males | Females |
| 2005 | 107 | 111 | 79 | 87 | 108 | 114 | 52 | 52 |
| 2006 | 102 | 113 | 48 | 44 | 102 | 114 | 48 | 44 |
| 2007 | 52 | 53 | 65 | 72 | 53 | 54 | 65 | 72 |
| 2008 | 104 | 109 | 79 | 79 | 104 | 112 | 79 | 79 |
| 2009 | 109 | 113 | 38 | 38 | 108 | 111 | 38 | 37 |
| 2010 | 138 | 156 | 64 | 68 | 138 | 156 | 64 | 68 |
| 2011 | 137 | 139 | 83 | 88 | 137 | 139 | 82 | 86 |
| 2012 | 127 | 131 | 66 | 88 | 126 | 136 | 66 | 88 |
| 2013 | 124 | 130 | 50 | 57 | 127 | 133 | 50 | 57 |
| 2014 | 137 | 142 | 89 | 95 | 137 | 143 | 88 | 93 |
| 2015 | 126 | 140 | 71 | 82 | 126 | 139 | 72 | 82 |
| 2016 | 162 | 176 | 78 | 80 | 162 | 176 | 78 | 81 |
| 2017 | 166 | 179 | 72 | 79 | 167 | 180 | 72 | 79 |
| 2018 | 129 | 150 | 50 | 55 | 129 | 152 | 50 | 57 |
| 2019 | 147 | 158 | 59 | 65 | 148 | 159 | 61 | 67 |
| **Total** | **1867** | **2000** | **991** | **1077** | **1872** | **2018** | **965** | **1042** |

Table S1.2: Number of individuals per number of observations, for each trait, sex and population.

| | Blue crown colouration | | | | Yellow breast patch colouration | | | |
|---|---|---|---|---|---|---|---|---|
| | Corsica | | Rouviere | | Corsica | | Rouviere | |
| N. Obs. | Males | Females | Males | Females | Males | Females | Males | Females |
| 1 | 774 | 882 | 419 | 494 | 778 | 824 | 396 | 470 |
| 2 | 275 | 300 | 139 | 124 | 276 | 297 | 137 | 124 |
| 3 | 104 | 102 | 56 | 59 | 105 | 107 | 55 | 57 |
| 4 | 35 | 34 | 25 | 25 | 34 | 37 | 26 | 25 |
| 5 | 12 | 20 | 4 | 8 | 12 | 19 | 4 | 7 |
| 6 | 4 | 6 | 1 | 3 | 4 | 6 | 1 | 3 |
| 7 | 1 | - | - | - | 1 | - | - | - |

Table S1.3: Repeatability of the colour measurements in each year and study site. D-Muro and E-Muro study sites were pooled together into a single category called Muro.

| | | Blue crown colouration | | Yellow breast patch colouration | |
|---|---|---|---|---|---|
| Study site | Year | Blue brightness | Blue UV chroma | Yellow brightness | Yellow chroma |
| D-Rouvière | 2005 | 0.802 | 0.815 | 0.543 | 0.752 |
| | 2006 | 0.681 | 0.807 | 0.602 | 0.684 |
| | 2007 | 0.695 | 0.814 | 0.526 | 0.823 |
| | 2008 | 0.763 | 0.833 | 0.404 | 0.664 |
| | 2009 | 0.752 | 0.801 | 0.580 | 0.767 |
| | 2010 | 0.918 | 0.871 | 0.819 | 0.720 |
| | 2011 | 0.737 | 0.800 | 0.466 | 0.671 |
| | 2012 | 0.749 | 0.781 | 0.504 | 0.612 |
| | 2013 | 0.751 | 0.770 | 0.605 | 0.580 |
| | 2014 | 0.920 | 0.896 | 0.805 | 0.821 |
| | 2015 | 0.916 | 0.842 | 0.728 | 0.746 |
| | 2016 | 0.923 | 0.831 | 0.846 | 0.709 |
| | 2017 | 0.946 | 0.892 | 0.765 | 0.681 |
| | 2018 | 0.934 | 0.849 | 0.811 | 0.799 |
| | 2019 | 0.873 | 0.886 | 0.74 | 0.66 |
| Muro | 2005 | 0.843 | 0.848 | 0.748 | 0.675 |
| | 2006 | 0.801 | 0.845 | 0.714 | 0.612 |
| | 2008 | 0.829 | 0.793 | 0.660 | 0.731 |
| | 2009 | 0.843 | 0.883 | 0.699 | 0.681 |
| | 2010 | 0.924 | 0.905 | 0.853 | 0.732 |
| | 2011 | 0.755 | 0.861 | 0.712 | 0.727 |
| | 2012 | 0.771 | 0.831 | 0.625 | 0.736 |
| | 2013 | 0.747 | 0.730 | 0.567 | 0.689 |
| | 2014 | 0.922 | 0.886 | 0.822 | 0.738 |
| | 2015 | 0.779 | 0.736 | 0.563 | 0.670 |
| | 2016 | 0.872 | 0.859 | 0.738 | 0.656 |
| | 2017 | 0.938 | 0.878 | 0.904 | 0.720 |
| | 2018 | 0.939 | 0.858 | 0.851 | 0.696 |
| | 2019 | 0.863 | 0.870 | 0.758 | 0.659 |
| E-Pirio | 2005 | 0.798 | 0.842 | 0.562 | 0.644 |
| | 2006 | 0.729 | 0.777 | 0.646 | 0.772 |
| | 2007 | 0.695 | 0.798 | 0.509 | 0.687 |
| | 2008 | 0.773 | 0.686 | 0.671 | 0.577 |
| | 2009 | 0.854 | 0.735 | 0.633 | 0.719 |
| | 2010 | 0.818 | 0.835 | 0.772 | 0.611 |
| | 2011 | 0.693 | 0.768 | 0.585 | 0.721 |
| | 2012 | 0.810 | 0.806 | 0.492 | 0.684 |
| | 2013 | 0.762 | 0.869 | 0.854 | 0.741 |
| | 2014 | 0.920 | 0.862 | 0.683 | 0.703 |
| | 2015 | 0.821 | 0.635 | 0.649 | 0.567 |
| | 2016 | 0.878 | 0.845 | 0.732 | 0.646 |
| | 2017 | 0.909 | 0.872 | 0.839 | 0.669 |
| | 2018 | 0.956 | 0.900 | 0.882 | 0.735 |
| | 2019 | 0.853 | 0.841 | 0.722 | 0.753 |

**SM-2: Variance Inflation Factors (VIFs) of the explanatory in the models associating the coloured traits and the climatic variables.**

Table S2.1. Variance inflation factors (VIF) of the fixed effects included in the models associating the colour with the climatic variables. Values for the excluded variables refer to the step before their exclusion (E. Seq.).

| | Corsica | | Rouviere | |
|---|---|---|---|---|
| **Variable** | **VIF** | **E. Seq.** | **VIF** | **E. Seq.** |
| Year (as continuous) | 3.50 | 1 | 1.12 | |
| Avg. temp. | 1.07 | | 1.16 | |
| Avg. prec. | 1.07 | | 1.06 | |

## SM-3: Pedigree information

The pedigree was constructed by assigning all fledged offspring in the two populations to their observed parents. This social approximation is a good proxy of the genetic pedigree as low levels of extrapair paternities have been detected in our populations. For the quantitative genetic analyses, we pruned the pedigree to retain the individuals used in each analysis. The information of each pedigree was obtained using the R package *pedantics* (Morrissey and Wilson 2010; see Table S3.1).

Table S3.1: Information about the pedigrees used in the animal models. The values represent the information of the pedigree including the fake identities created to maintain the sibship information. The values in italics and in brackets represents the information of the pedigree not including the fake identities.

| | Corsica | | | | Rouviere | | | |
|---|---|---|---|---|---|---|---|---|
| | Blue Colours Males | Blue Colours Females | Yellow Colours Males | Yellow Colours Females | Blue Colours Males | Blue Colours Females | Yellow Colours Males | Yellow Colours Females |
| Records | 1856 [*1787*] | 1758 [*1698*] | 1864 [*1795*] | 1767 [*1707*] | 1148 [*1103*] | 1155 [*1123*] | 1129 [*1085*] | 1138 [*1107*] |
| Maximum depth | 16 [*16*] | 15 [*15*] | 16 [*16*] | 15 [*15*] | 16 [*16*] | 16 [*16*] | 16 [*16*] | 16 [*16*] |
| Maternities | 694 [*661*] | 486 [*460*] | 696 [*663*] | 489 [*462*] | 618 [*603*] | 497 [*485*] | 605 [*590*] | 493 [*481*] |
| Paternities | 694 [*646*] | 486 [*443*] | 696 [*648*] | 489 [*446*] | 618 [*577*] | 497 [*467*] | 605 [*565*] | 493 [*464*] |
| Full sibs | 152 [*140*] | 86 [*77*] | 152 [*140*] | 87 [*77*] | 244 [*234*] | 155 [*144*] | 235 [*225*] | 154 [*143*] |
| Maternal sibs | 277 [*271*] | 157 [*154*] | 277 [*271*] | 160 [*156*] | 379 [*375*] | 240 [*236*] | 367 [*363*] | 238 [*234*] |
| Paternal sibs | 282 [*276*] | 140 [*133*] | 282 [*276*] | 142 [*135*] | 363 [*356*] | 231 [*223*] | 345 [*338*] | 226 [*218*] |
| Pairwise relatedness | 0.0008 [*0.0008*] | 0.0006 [*0.0006*] | 0.0008 [*0.0008*] | 0.0006 [*0.0006*] | 0.003 [*0.003*] | 0.002 [*0.002*] | 0.003 [*0.003*] | 0.002 [*0.002*] |

## SM-4: Alternative models

In order to check that the quantitative genetic parameters reported from the animal models were not an artefact generated by the chosen prior we re-ran the animal models with two alternative priors (see below). The results of the heritabilities drawn from the models using these priors showed that the reported heritability values and 95% Credible Intervals (C.I.) were similar (see Tables S4.1, S4.2 and S4.3). The significance of the heritabilities drawn for the animal model were obtained by comparing the Deviance Information Criterion (DIC) of the model with and without the pedigree information. A heritability was considered significant when the model with the pedigree had a DIC <2 than the model without the pedigree.

**Prior 1 (used in the analyses presented in the main text):**
$G_A$: V = 1, nu = 1, alpha.mu=0, alpha.V= 1000
$G_{PE}$: V = 1, nu = 1, alpha.mu=0, alpha.V= 1000
$G_{YR}$: V = 1, nu = 1, alpha.mu=0, alpha.V= 1000
R: V = 1, nu = 0.002

**Prior 2:**
$G_A$: V = 1, nu = 2, alpha.mu=0, alpha.V= 1000
$G_{PE}$: V = 1, nu = 2, alpha.mu=0, alpha.V= 1000
$G_{YR}$: V = 1, nu = 2, alpha.mu=0, alpha.V = 1000
R: V = 1, nu = 0.002

**Prior 3:**
$G_A$: V = 1, nu = 3, alpha.mu=0, alpha.V= 1000
$G_{PE}$: V = 1, nu = 3, alpha.mu=0, alpha.V= 1000
$G_{YR}$: V = 1, nu = 3, alpha.mu=0, alpha.V = 1000
R: V = 1, nu = 0.002

Table S4.1: Heritability ($V_A/V_P$) and 95% Credible Intervals (CI) obtained from the animal models of each coloured trait in each sex and population, using Prior 1.

| Prior1 | | | | | | | |
|---|---|---|---|---|---|---|---|
| Corsica | | | | | | | |
| | Males | | | | Females | | |
| | *h2* | 95% C.I. | Sig. | | *h2* | 95% C.I. | Sig |
| Blue crown UV chroma | 0.167 | [0.088, 0.237] | YES | | 0.161 | [0.086, 0.233] | YES |
| Blue crown brightness | 0.029 | [0.000001, 0.070] | YES | | 0.053 | [0.00002, 0.103] | YES |
| Yellow breast patch chroma | 0.074 | [0.0000001, 0.143] | YES | | 0.026 | [0.00000003, 0.074] | NO |
| Yellow breast patch brightness | 0.078 | [0.00001, 0.134] | YES | | 0.034 | [0.00000005, 0.079] | YES |
| D-Rouvière | | | | | | | |
| | Males | | | | Females | | |
| Blue crown UV chroma | 0.058 | [0.00000005, 0.128] | YES | | 0.173 | [0.063, 0.266] | YES |
| Blue crown brightness | 0.068 | [0.000000002, 0.135] | YES | | 0.073 | [0.0000003, 0.132] | YES |
| Yellow breast patch chroma | 0.110 | [0.016, 0.193] | YES | | 0.080 | [0.00001, 0.151] | YES |
| Yellow breast patch brightness | 0.039 | [0.0000002, 0.097] | YES | | 0.029 | [0.00000009, 0.088] | NO |

Table S4.2: Heritability ($V_A/V_P$) and 95% Credible Intervals (CI) obtained from the animal models of each coloured trait in each sex and population, using Prior 2.

| Prior 2 | | | | | | | |
|---|---|---|---|---|---|---|---|
| Corsica | | | | | | | |
| | Males | | | | Females | | |
| | *h2* | 95% C.I. | Sig. | | *h2* | 95% C.I. | Sig |
| Blue crown UV chroma | 0.166 | [0.092, 0.241] | YES | | 0.163 | [0.089, 0.244] | YES |
| Blue crown brightness | 0.030 | [0.0000004, 0.070] | YES | | 0.052 | [0.00003, 0.103] | YES |
| Yellow breast patch chroma | 0.073 | [0.0000001, 0.130] | YES | | 0.025 | 0.0000000007, 0.076] | NO |
| Yellow breast patch brightness | 0.077 | [0.000003, 0.138] | YES | | 0.034 | [0.0000001, 0.078] | YES |
| D-Rouviére | | | | | | | |
| | Males | | | | Females | | |
| Blue crown UV chroma | 0.056 | [0.00006, 0.119] | YES | | 0.172 | [0.066, 0.265] | YES |
| Blue crown brightness | 0.066 | [0.0000005, 0.129] | YES | | 0.071 | [0.00001, 0.129] | YES |
| Yellow breast patch chroma | 0.111 | [0.017, 0.204] | YES | | 0.082 | [0.000003, 0.155] | YES |
| Yellow breast patch brightness | 0.040 | [0.000000007, 0.095] | YES | | 0.028 | [0.00000002, 0.088] | NO |

Table S4.3: Heritability ($V_A/V_P$) and 95% Credible Intervals (CI) obtained from the animal models of each coloured trait in each sex and population, using Prior 3.

| Prior 3 | | | | | | | |
|---|---|---|---|---|---|---|---|
| Corsica | | | | | | | |
| | Males | | | | Females | | |
| | *h2* | 95% C.I. | Sig. | | *h2* | 95% C.I. | Sig |
| Blue crown UV chroma | 0.167 | [0.093, 0.238] | YES | | 0.161 | [0.083, 0.233] | YES |
| Blue crown brightness | 0.031 | [0.000000006, 0.073] | YES | | 0.051 | [0.00000003, 0.102] | YES |
| Yellow breast patch chroma | 0.075 | [0.00003, 0.138] | YES | | 0.026 | [0.00000008, 0.076] | NO |
| Yellow breast patch brightness | 0.077 | [0.00014, 0.138] | YES | | 0.033 | [0.0000001, 0.079] | YES |
| D-Rouviére | | | | | | | |
| | Males | | | | Females | | |
| Blue crown UV chroma | 0.057 | [0.00001, 0.125] | YES | | 0.176 | [0.072, 0.130] | YES |
| Blue crown brightness | 0.069 | [0.0000001, 0.130] | YES | | 0.074 | [0.000009, 0.131] | YES |
| Yellow breast patch chroma | 0.107 | [0.008, 0.188] | YES | | 0.080 | [0.00001, 0.157] | YES |
| Yellow breast patch brightness | 0.036 | [0.0000001, 0.099] | YES | | 0.026 | [0.00000001, 0.086] | NO |

## SM-5 Results of the temporal trends in the breeding values obtained from less conservative models

Table S5.1: Posterior mode estimates and CI (95%) for the linear regression of the BLUPs obtained from the animal models not including year (as a continuous variable) as a fixed effect, against the mean year.

| **Corsica** | | | | |
|---|---|---|---|---|
| | Males | | Females | |
| | Estimate | 95% CI | Estimate | 95% CI |
| Blue crown UV chroma | -0.0022 | [-0.0043, 0.0011] | -0.0016 | [-0.0036, 0.0011] |
| Blue crown brightness | 0.0002 | [-0.196, 0.169] | 0.0019 | [-0.236, 0.131] |
| Yellow breast patch chroma | -0.00007 | [-0.009, 0.006] | 0.00001 | [-0.0042, 0.0039] |
| Yellow breast patch brightness | 0.0027 | [-0.192, 0.164] | 0.0012 | [-0.1204, 0.1181] |
| **D-Rouvière** | | | | |
| | Males | | Females | |
| Blue crown UV chroma | -0.0002 | [-0.0035, 0.002] | -0.0021 | [-0.0063, 0.0013] |
| Blue crown brightness | -0.119 | [-0.685, 0.382] | -0.067 | [-0.539, 0.280] |
| Yellow breast patch chroma | -0.0009 | [-0.019, 0.020] | -0.00001 | [-0.018, 0.009] |
| Yellow breast patch brightness | -0.005 | [-0.198, 0.278] | 0.0014 | [-0.197, 0.222] |

## SM-6 Results of the quantitative genetic models including the maternal effects

To explore the effects of the maternal effects on the heritabilities and trends in the breeding values reported in the main manuscript we reran all the animal models with the same structure than in the main text but including the ring of the mother as a random factor and keeping a fake mother identity for those individuals with unknown mothers, as in the main analyses.

Our results show that the variance explained by the maternal effects is low (range: 0.008-0.048, mean: 0.022; Table S6.1). Also, the differences of the heritabilities obtained from the models including the maternal effects and those presented in the main text are very small (range: 0.004 – 0.027, mean=0.011), and the confidence intervals overlap (see Table S6.2). Finally, including the maternal effects in the animal models did not change our estimates of the temporal trends in the breeding values (Table S6.3).

Table S6.1: Variance explained by the maternal effects ($V_M/V_P$) and 95% confidence intervals, in Corsica and D-Rouvière.

| | | | | |
|---|---|---|---|---|
| **Corsica** | | | | |
| | Males | | Females | |
| | *Mat. Eff.* | 95% CI | *Mat. Eff.* | 95% CI |
| Blue crown UV chroma | 0.008 | [<0.001, 0.030] | 0.011 | [<0.001, 0.040] |
| Blue crown brightness | 0.017 | [<0.001, 0.056] | 0.014 | [<0.001, 0.048] |
| Yellow breast patch chroma | 0.022 | [<0.001, 0.073] | 0.022 | [<0.001, 0.070] |
| Yellow breast patch brightness | 0.016 | [<0.001, 0.058] | 0.010 | [<0.001, 0.038] |
| **D-Rouvière** | | | | |
| | Males | | Females | |
| Blue crown UV chroma | 0.019 | [<0.001, 0.064] | 0.043 | [<0.001, 0.126] |
| Blue crown brightness | 0.048 | [<0.001, 0.115] | 0.022 | [<0.001, 0.076] |
| Yellow breast patch chroma | 0.016 | [<0.001, 0.057] | 0.028 | [<0.001, 0.089] |
| Yellow breast patch brightness | 0.020 | [<0.001, 0.067] | 0.041 | [<0.001, 0.109] |

Table S6.2: Heritability ($h^2$; $V_A/V_P$) and 95% confidence intervals (CI) present in the main text (Without Mat. Eff.) and from the models including the maternal effects (With Mat. Eff.) for Corsica and Rouviere.

| | | | | | | | | |
|---|---|---|---|---|---|---|---|---|
| **Corsica** | | | | | | | | |
| | Males | | | | Females | | | |
| | Without Mat. Eff. | | With Mat. Eff. | | Without Mat. Eff. | | With Mat. Eff. | |
| | *$h^2$* | 95% CI | *$h^2$* | 95% CI | *$h^2$* | 95% CI | *$h^2$* | 95% CI |
| Blue crown UV chroma | 0.167 | [0.088, 0.237] | 0.162 | [0.085, 0.235] | 0.161 | [0.086, 0.233] | 0.153 | [0.075, 0.228] |
| Blue crown brightness | 0.029 | [<0.001, 0.070] | 0.025 | [<0.001, 0.064] | 0.053 | [<0.001, 0.103] | 0.042 | [<0.001, 0.092] |
| Yellow breast patch chroma | 0.074 | [<0.001, 0.143] | 0.058 | [<0.001, 0.122] | 0.026 | [<0.001, 0.074] | 0.021 | [<0.001, 0.068] |
| Yellow breast patch brightness | 0.078 | [<0.001, 0.134] | 0.067 | [<0.001, 0.126] | 0.034 | [<0.001, 0.079] | 0.028 | [<0.001, 0.073] |
| **D-Rouvière** | | | | | | | | |
| | Males | | | | Females | | | |
| Blue crown UV chroma | 0.058 | [<0.001, 0.128] | 0.051 | [<0.001, 0.112] | 0.173 | [0.063, 0.266] | 0.146 | [0.043, 0.261] |
| Blue crown brightness | 0.068 | [<0.001, 0.135] | 0.042 | [<0.001, 0.103] | 0.073 | [<0.001, 0.132] | 0.058 | [<0.001, 0.122] |
| Yellow breast patch chroma | 0.110 | [0.016, 0.193] | 0.101 | [<0.001, 0.186] | 0.080 | [<0.001, 0.151] | 0.062 | [<0.001, 0.136] |
| Yellow breast patch brightness | 0.039 | [<0.001, 0.097] | 0.029 | [<0.001, 0.079] | 0.029 | [<0.001, 0.088] | 0.021 | [<0.001, 0.069] |

Table S6.3: Temporal trends in the breeding values and 95% confidence intervals comparing results from the main text (without Mat. Eff.) and those obtained from models including the maternal effects (with Matt. Eff.). BUVC stands for blue crown UV chroma, BB stands for blue brightness, YC stands for yellow breast patch chroma and YB stands for yellow brightness.

| **Corsica** | | | | | | | | |
|---|---|---|---|---|---|---|---|---|
| | Males | | | | Females | | | |
| | Without Mat. Eff. | | With Matt. Eff. | | Without Mat. Eff. | | With Matt. Eff. | |
| | *Est* | 95% CI | *Est* | 95% CI | *Est* | 95% CI | *Est* | 95% CI |
| BUVC | -0.0014 | [-0.0043, 0.0012] | -0.0019 | [-0.0041, 0.0013] | -0.0002 | [-0.0036, 0.0010] | -0.0016 | [-0.0032, 0.0013] |
| BB | 0.0107 | [-0.2155, 0.1548] | 0.0005 | [-0.1637, 0.1793] | -0.0012 | [-0.2429, 0.1255] | -0.0018 | [-0.2050, 0.1214] |
| YC | -0.00005 | [-0.0100, 0.0060] | 0.0000006 | [-0.007, 0.0007] | -0.00002 | [-0.0043, 0.0044] | 0.00002 | [-0.0037, 0.0042] |
| YB | 0.0043 | [-0.1738, 0.1791] | -0.0005 | [-0.1577, 0.1560] | -0.0004 | [-0.1306, 0.099] | 0.00006 | [-0.1225, 0.0967] |
| **D-Rouvière** | | | | | | | | |
| | Males | | | | Females | | | |
| BUVC | -0.00003 | [-0.0003, 0.0023] | -0.00007 | [-0.0035, 0.0018] | -0.0010 | [-0.0055; 0.0019] | -0.00008 | [-0.0052, 0.0018] |
| BB | -0.00009 | [-0.6115, 0.3825] | -0.0046 | [-0.4470, 0.2846] | 0.0332 | [-0.5408; 0.2943] | -0.0014 | [-0.4358, 0.3348] |
| YC | -0.0011 | [-0.0208, 0.0180] | -0.0027 | [-0.0176, 0.0197] | -0.0007 | [-0.0156; 0.0125] | -0.00005 | [-0.0134, 0.0128] |
| YB | -0.0012 | [-0.1591, 0.3134] | 0.0002 | [-0.1328, 0.2920] | 0.0009 | [-0.1935; 0.2288] | -0.0005 | [-0.1461, 0.2050] |

In addition, we also ran the models removing the observations of those individuals with unknown mothers. Although this led to a substantial reduction of the available data because in our dataset around 60% of the observations have an unknown mother, because they were not born in a nest-box (Blue colours: 3664/5935; Yellow colours: 3651/5897), the results are very similar to those presented with the full dataset including fake identities for the unknown mothers (see Tables S6.4-S6.5).

Table S6.4 Heritability ($h^2$; $V_A/V_P$) and 95% confidence intervals (CI) of the models including the maternal effects and the individuals with known mothers in Corsica

| Corsica | | | | |
|---|---|---|---|---|
| | Males | | Females | |
| | $h^2$ | **95% C.I.** | $h^2$ | **95% C.I.** |
| BB | 0.030 | [<0.001, 0.086] | 0.039 | [<0.001, 0.117] |
| BUVC | 0.136 | [0.053, 0.227] | 0.121 | [<0.001, 0.264] |
| YB | 0.069 | [<0.001, 0.159] | 0.037 | [<0.001, 0.111] |
| YC | 0.045 | [<0.001, 0.115] | 0.025 | [<0.001, 0.086] |

Table S6.5 Heritability ($h^2$; $V_A/V_P$) and 95% confidence intervals (CI) of the models including the maternal effects and the individuals with known mothers in D-Rouvière.

| Rouviere | | | | |
|---|---|---|---|---|
| | Males | | Females | |
| | $h^2$ | **95% C.I.** | $h^2$ | **95% C.I.** |
| BB | 0.022 | [<0.001, 0.073] | 0.035 | [<0.001, 0.108] |
| BUVC | 0.031 | [<0.001, 0.085] | 0.074 | [<0.001, 0.195] |
| YB | 0.020 | [<0.001, 0.067] | 0.017 | [<0.001, 0.065] |
| YC | 0.068 | [<0.001, 0.138] | 0.157 | [<0.001, 0.300] |

## SM-7 Results of the models including age (1-year-old vs ≥2-year-old) as a covariate.

Table S7.1. Linear temporal trends for blue crown and yellow breast patch colourations of the blue tits in Corsica and D-Rouvière including age as a covariate. Significant ($p<0.05$) variables are in bold. $R^2_{cond}$ represents the variance explained by both the fixed and random effects included in the model, $R^2_{mar}$ represents the variance explained by the fixed factors alone.

| | **Corsica** | | | | **D-Rouvière** | | | |
|---|---|---|---|---|---|---|---|---|
| **Blue crown UV chroma (n. obs.: Corsica=3867; D-Rouvière=2068)** | | | | | | | | |
| **Fixed effects** | Est | SE | F | P | Est | SE | F | P |
| Year | **-0.004** | **0.001** | **$F_{1,13.0}$=13.369** | **0.002** | **-0.002** | **0.001** | **$F_{1,13.05}$=5.632** | **0.033** |
| Sex(fem) | **-2.010** | **0.392** | **$F_{1,2796.0}$=26.178** | **<0.001** | **-1.257** | **0.543** | **$F_{1,1405.24}$=5.362** | **0.020** |
| Age(adults) | **0.011** | **0.0007** | **$F_{1,3607.6}$=213.233** | **<0.001** | **0.010** | **0.001** | **$F_{1,1793.37}$=106.866** | **<0.001** |
| Year*sex(fem) | **0.0009** | **0.0001** | **$F_{1,2795.9}$=25.278** | **<0.001** | **0.0006** | **0.0002** | **$F_{1,1404.51}$=5.046** | **0.024** |
| | $R^2_{cond}$: 0.765; $R^2_{mar}$: 0.388 | | | | $R^2_{cond}$: 0.720; $R^2_{mar}$: 0.357 | | | |
| **Blue crown brightness (n. obs.: Corsica=3867; D-Rouvière=2068)** | | | | | | | | |
| **Fixed effects** | Est | SE | F | P | Est | SE | F | P |
| Year | -0.181 | 0.136 | $F_{1,13.1}$=1.349 | 0.266 | **-0.384** | **0.164** | **$F_{1,13.01}$=6.155** | **0.027** |
| Sex (fem) | -96.872 | 55.626 | $F_{1,2775.1}$=3.032 | 0.081 | 77.399 | 88.303 | $F_{1,1331.55}$=0.768 | 0.380 |
| Age(adults) | **0.870** | **0.118** | **$F_{1,3769.9}$=54.322** | **<0.001** | **1.139** | **0.177** | **$F_{1,1919.76}$=41.304** | **<0.001** |
| Year*sex(fem) | 0.046 | 0.027 | $F_{1,2774.9}$=2.887 | 0.089 | -0.039 | 0.043 | $F_{1,1330.52}$=0.813 | 0.367 |
| | $R^2_{cond}$: 0.440; $R^2_{mar}$: 0.105 | | | | $R^2_{cond}$: 0.517; $R^2_{mar}$: 0.160 | | | |
| **Yellow breast patch chroma (n. obs.: Corsica=3890; D-Rouvière=2007)** | | | | | | | | |
| **Fixed effects** | Est | SE | F | P | Est | SE | F | P |
| Year | **-0.012** | **0.002** | **$F_{1,13.2}$=13.957** | **0.003** | -0.009 | 0.005 | $F_{1,13.08}$=3.123 | 0.100 |
| Sex (fem) | **-9.805** | **1.935** | **$F_{1,2737.0}$=25.665** | **<0.001** | -0.731 | 3.006 | $F_{1,1244.99}$=0.059 | 0.807 |
| Age(adults) | **0.011** | **0.004** | **$F_{1,3790.3}$=7.558** | **0.006** | **-0.022** | **0.005** | **$F_{1,1832.76}$=13.140** | **<0.001** |
| Year*sex(fem) | **0.004** | **0.0009** | **$F_{1,2736.7}$=25.308** | **<0.001** | 0.0003 | 0.001 | $F_{1,1244.14}$=0.056 | 0.811 |
| | $R^2_{cond}$: 0.385; $R^2_{mar}$: 0.140 | | | | $R^2_{cond}$: 0.480; $R^2_{mar}$: 0.067 | | | |
| **Yellow breast patch brightness (n. obs.: Corsica=3890; D-Rouvière=2007)** | | | | | | | | |
| **Fixed effects** | Est | SE | F | P | Est | SE | F | P |
| Year | **-0.306** | **0.074** | **$F_{1,13.1}$=21.387** | **<0.001** | **-0.350** | **0.116** | **$F_{1,13.09}$=9.402** | **0.008** |
| Sex(fem) | **138.326** | **42.509** | **$F_{1,2601.3}$=10.589** | **0.001** | 7.780 | 61.230 | $F_{1,1167.07}$=0.016 | 0.898 |
| Age(adults) | **-0.316** | **0.090** | **$F_{1,3784.4}$=12.121** | **<0.001** | **-0.496** | **0.122** | **$F_{1,1885.33}$=16.519** | **<0.001** |
| Year*sex(fem) | **-0.068** | **0.021** | **$F_{1,2601.0}$=10.494** | **0.001** | -0.003 | 0.030 | $F_{1,1166.04}$=0.014 | 0.904 |
| | $R^2_{cond}$: 0.393; $R^2_{mar}$: 0.199 | | | | $R^2_{cond}$: 0.512; $R^2_{mar}$: 0.174 | | | |

Table S7.2. Associations between the four colour components and the average temperature (Avg. temp.) and average precipitation (Avg. prec.) during moulting in Corsica and D-Rouvière, including age as a covariate. Significant ($p<0.05$) variables are in bold. $R^2_{cond}$ represents the variance explained by both the fixed and random effects included in the model, $R^2_{mar}$ represents the variance explained by the fixed factors alone. Year was not included in the Corsican models due to its high collinearity with the climatic variables (see Methods).

| | **Corsica** | | | | **D-Rouvière** | | | |
|---|---|---|---|---|---|---|---|---|
| **Blue crown UV chroma (n. obs.: Corsica=3867; D-Rouvière=2068)** | | | | | | | | |
| **Fixed Effects** | Est | SE | F | P | Est | SE | F | P |
| Avg. temp. | -0.023 | 0.011 | $F_{1,12.0}$=3.596 | 0.082 | -0.006 | 0.008 | $F_{1,10.99}$=0.348 | 0.566 |
| Avg. prec. | 0.020 | 0.013 | $F_{1,12.0}$=1.549 | 0.237 | -0.004 | 0.005 | $F_{1,10.98}$=0.599 | 0.455 |
| Sex(fem) | **-0.102** | **0.034** | **$F_{1,3659.8}$=8.957** | **0.002** | **-0.094** | **0.037** | **$F_{1,1977.82}$=6.204** | **0.012** |
| Age(adults) | **0.011** | **0.0007** | **$F_{1,3605.4}$=215.05** | **<0.001** | **0.010** | **0.001** | **$F_{1,1794.16}$=106.44** | **<0.001** |
| Year | | | | | -0.002 | 0.001 | $F_{1,11.02}$=4.189 | 0.065 |
| Avg. temp.*sex(fem) | **0.003** | **0.001** | **$F_{1,3692.4}$=4.642** | **0.031** | 0.002 | 0.001 | $F_{1,1968.35}$=2.184 | 0.139 |
| Avg. prec.*sex(fem) | **-0.007** | **0.002** | **$F_{1,3838.8}$=14.192** | **<0.001** | 0.0009 | 0.001 | $F_{1,2030.65}$=0.604 | 0.437 |
| | $R^2_{cond}$: 0.766; $R^2_{mar}$:0.328 | | | | $R^2_{cond}$: 0.731; $R^2_{mar}$:0.357 | | | |
| **Blue crown brightness (n. obs.: Corsica=3867; D-Rouvière=2068)** | | | | | | | | |
| **Fixed Effects** | Est | SE | F | P | Est | SE | F | P |
| Avg. temp. | -1.415 | 1.230 | $F_{1,12.0}$=0.998 | 0.337 | 2.307 | 1.108 | $F_{1,10.97}$=4.723 | 0.052 |
| Avg. prec. | 0.185 | 1.431 | $F_{1,12.1}$=0.194 | 0.891 | 0.785 | 0.728 | $F_{1,10.95}$=1.129 | 0.310 |
| Sex(fem) | **-11.257** | **5.172** | **$F_{1,3820.8}$=4.736** | **0.029** | -5.537 | 6.481 | $F_{1,2035.43}$=0.729 | 0.393 |
| Age(adults) | **0.868** | **0.118** | **$F_{1,3769.9}$=50.004** | **<0.001** | **1.141** | **0.177** | **$F_{1,1920.15}$=41.386** | **<0.001** |
| Year | | | | | -0.489 | 0.156 | $F_{1,11.0}$=9.726 | 0.009 |
| Avg. temp.*sex(fem) | 0.381 | 0.219 | $F_{1,3830.2}$=3.013 | 0.082 | 0.157 | 0.298 | $F_{1,2031.43}$=0.280 | 0.596 |
| Avg. prec.*sex(fem) | 0.024 | 0.277 | $F_{1,3790.9}$=0.007 | 0.930 | -0.039 | 0.200 | $F_{1,2046.35}$=0.037 | 0.846 |
| | $R^2_{cond}$: 0.452; $R^2_{mar}$: 0.103 | | | | $R^2_{cond}$: 0.528; $R^2_{mar}$:0.231 | | | |
| **Yellow breast patch chroma (n. obs.: Corsica=3890; D-Rouvière=2007)** | | | | | | | | |
| **Fixed Effects** | Est | SE | F | P | Est | SE | F | P |
| Avg. temp. | **-0.072** | **0.025** | **$F_{1,12.0}$=5.580** | **0.035** | 0.085 | 0.037 | $F_{1,11.01}$=4.236 | 0.064 |
| Avg. prec. | 0.060 | 0.030 | $F_{1,12.1}$=2.034 | 0.179 | -0.011 | 0.024 | $F_{1,10.99}$=0.004 | 0.947 |
| Sex(fem) | **-0.617** | **0.179** | **$F_{1,3838.2}$=11.793** | **<0.001** | 0.038 | 0.020 | $F_{1,1967.72}$=3.408 | 0.065 |
| Age(adults) | **0.011** | **0.004** | **$F_{1,3787.9}$=7.969** | **0.004** | **-0.022** | **0.005** | **$F_{1,1833.46}$=14.88** | **<0.001** |
| Year | | | | | **-0.013** | **0.005** | **$F_{1,11.07}$=6.268** | **0.029** |
| Avg. temp.*sex(fem) | **0.024** | **0.007** | **$F_{1,3849.3}$=10.525** | **0.001** | **-0.020** | **0.009** | **$F_{1,1962.87}$=4.564** | **0.032** |
| Avg. prec. *sex(fem) | **-0.036** | **0.009** | **$F_{1,3815.6}$=14.118** | **<0.001** | **0.020** | **0.006** | **$F_{1,1987.99}$=9.279** | **0.002** |
| | $R^2_{cond}$: 0.395; $R^2_{mar}$: 0.135 | | | | $R^2_{cond}$: 0.499; $R^2_{mar}$: 0.141 | | | |
| **Yellow breast patch brightness (n. obs.: Corsica=3890; D-Rouvière=2007)** | | | | | | | | |
| **Fixed Effects** | Est | SE | F | P | Est | SE | F | P |
| Avg. temp. | **-1.996** | **0.777** | **$F_{1,12.0}$=7.612** | **0.017** | 0.263 | 0.910 | $F_{1,11.03}$=0.061 | 0.808 |
| Avg. prec. | 0.698 | 0.904 | $F_{1,12.1}$=1.309 | 0.274 | 0.660 | 0.598 | $F_{1,11.01}$=0.942 | 0.352 |
| Sex(fem) | 6.381 | 3.991 | $F_{1,3849.0}$=2.555 | 0.109 | 2.424 | 4.379 | $F_{1,1983.46}$=0.306 | 0.579 |
| Age(adults) | **-0.324** | **0.090** | **$F_{1,3783.8}$=12.695** | **<0.001** | **-0.499** | **0.122** | **$F_{1,1885.26}$=16.731** | **<0.001** |
| Year | | | | | **-0.338** | **0.129** | **$F_{1,11.05}$=6.857** | **0.023** |
| Avg. temp.*sex(fem) | -0.269 | 0.169 | $F_{1,3858.0}$=1.524 | 0.112 | -0.076 | 0.201 | $F_{1,1981.13}$=0.145 | 0.702 |
| Avg. prec.*sex(fem) | **0.658** | **0.213** | **$F_{1,3795.4}$=9.536** | **0.002** | -0.166 | 0.137 | $F_{1,1974.66}$=1.468 | 0.225 |
| | $R^2_{cond}$: 0.417; $R^2_{mar}$: 0.164 | | | | $R^2_{cond}$:0.528; $R^2_{mar}$: 0.177 | | | |

Table S7.3: Heritability ($V_A/V_P$) and 95% Credible Intervals (CI) obtained from the animal models of each coloured trait in each sex and population. The animal models included year (as a continuous variable), age (coded as 1-year-old vs ≥2-years-old), and site (in the Corsican models).

| Corsica | | | | | | | |
|---|---|---|---|---|---|---|---|
| | Males | | | | Females | | |
| | *h2* | 95% C.I. | Sig. | | *h2* | 95% C.I. | Sig |
| Blue crown UV chroma | 0.170 | [0.098, 0.239] | YES | | 0.159 | [0.078, 0.235] | YES |
| Blue crown brightness | 0.032 | [<0.001, 0.075] | YES | | 0.055 | [<0.001, 0.107] | YES |
| Yellow breast patch chroma | 0.076 | [<0.001, 0.137] | YES | | 0.025 | [<0.001,0.075] | NO |
| Yellow breast patch brightness | 0.073 | [<0.001, 0.130] | YES | | 0.032 | [<0.001, 0.079] | YES |
| D-Rouviére | | | | | | | |
| | Males | | | | Females | | |
| Blue crown UV chroma | 0.063 | [<0.001, 0.131] | YES | | 0.185 | [0.078, 0.283] | YES |
| Blue crown brightness | 0.068 | [<0.001, 0.132] | YES | | 0.084 | [0.018, 0.154] | YES |
| Yellow breast patch chroma | 0.113 | [0.014, 0.207] | YES | | 0.078 | [<0.001, 0.153] | YES |
| Yellow breast patch brightness | 0.038 | [<0.001, 0.091] | YES | | 0.029 | [<0.001, 0.090] | NO |

Table S7.4: Posterior mode estimates and CI (95%) of the linear regression of the BLUPs obtained from the animal models against the mean year. The animal models from which these BLUPs were obtained included year (as a continuous variable), age (coded as coded as 1-year-old vs ≥2-years-old), and site (in the Corsican models)

| **Corsica** | | | | |
|---|---|---|---|---|
| | Males | | Females | |
| | Estimate | 95% CI | Estimate | 95% CI |
| Blue crown UV chroma | -0.0022 | [-0.0043, 0.0011] | -0.0016 | [-0.0036, 0.0011] |
| Blue crown brightness | 0.0002 | [-0.196, 0.169] | 0.0019 | [-0.236, 0.131] |
| Yellow breast patch chroma | -0.00007 | [-0.009, 0.006] | 0.00001 | [-0.0042, 0.0039] |
| Yellow breast patch brightness | 0.0027 | [-0.192, 0.164] | 0.0012 | [-0.1204, 0.1181] |
| **D-Rouvière** | | | | |
| | Males | | Females | |
| Blue crown UV chroma | -0.0002 | [-0.0035, 0.002] | -0.0021 | [-0.0063, 0.0013] |
| Blue crown brightness | -0.119 | [-0.685, 0.382] | -0.067 | [-0.539, 0.280] |
| Yellow breast patch chroma | -0.0009 | [-0.019, 0.020] | -0.00001 | [-0.018, 0.009] |
| Yellow breast patch brightness | -0.005 | [-0.198, 0.278] | 0.0014 | [-0.197, 0.222] |

## SM-8: Information about the fixed and random effects of the models and results of the models with unscaled predictors

Table S8.1. Linear temporal trends for blue crown and yellow breast patch colourations of the blue tits in Corsica and D-Rouvière. Significant ($p<0.05$) variables are in bold. $R^2_{cond}$ represents the variance explained by both the fixed and random effects included in the model, $R^2_{mar}$ represents the variance explained by the fixed factors alone. Predictors were scaled to mean of zero and standard deviation of unity.

| | **Corsica** | | | | **D-Rouvière** | | | |
|---|---|---|---|---|---|---|---|---|
| **Blue crown UV chroma (n. obs.: Corsica=3867; D-Rouvière=2068)** | | | | | | | | |
| **Fixed effects** | Est | SE | F | P | Est | SE | F | P |
| Year | **-0.019** | **0.004** | **$F_{1,13.03}$=12.524** | **0.003** | **-0.011** | **0.004** | **$F_{1,13.03}$=5.038** | **0.042** |
| Sex(fem) | **-0.035** | **0.0008** | **$F_{1,2213.7}$=1614.7** | **<0.001** | **-0.038** | **0.001** | **$F_{1,1169.4}$=1024.9** | **<0.001** |
| Year*sex(fem) | **0.004** | **0.0008** | **$F_{1,2790.5}$=23.383** | **<0.001** | **0.002** | **0.001** | **$F_{1,1407.8}$=3.927** | **0.047** |
| **Random effects** | Variance | Std. Dev. | | | Variance | Std. Dev. | | |
| Individual | 0.0001 | 0.012 | | | 0.0001 | 0.012 | | |
| Year | 0.0003 | 0.019 | | | 0.0003 | 0.018 | | |
| Site | 0.00008 | 0.009 | | | - | - | | |
| Residual | 0.0004 | 0.020 | | | 0.0004 | 0.020 | | |
| | $R^2_{cond}$:0.740; $R^2_{mar}$: 0.365 | | | | $R^2_{cond}$:0.696; $R^2_{mar}$: 0.338 | | | |
| **Blue crown brightness (n. obs.: Corsica=3867; D-Rouvière=2068)** | | | | | | | | |
| **Fixed effects** | Est | SE | F | P | Est | SE | F | P |
| Year | -0.756 | 0.586 | $F_{1,13.06}$=1.287 | 0.276 | **-1.557** | **0.701** | **$F_{1,13.01}$=5.670** | **0.033** |
| Sex (fem) | **-2.378** | **0.118** | **$F_{1,2125.6}$=399.71** | **<0.001** | **-2.332** | **0.189** | **$F_{1,1097.0}$=150.75** | **<0.001** |
| Year*sex(fem) | 0.190 | 0.117 | $F_{1,2774.0}$=2.653 | 0.103 | -0.195 | 0.187 | $F_{1,1337.7}$=1.093 | 0.295 |
| **Random effects** | Variance | Std. Dev. | | | Variance | Std. Dev. | | |
| Individual | 1.261 | 1.123 | | | 2.487 | 1.577 | | |
| Year | 5.294 | 2.301 | | | 7.383 | 2.717 | | |
| Site | 0.055 | 0.234 | | | - | - | | |
| Residual | 11.178 | 3.343 | | | 13.890 | 3.727 | | |
| | $R^2_{cond}$:0.431; $R^2_{mar}$: 0.094 | | | | $R^2_{cond}$: 0.502; $R^2_{mar}$: 0.148 | | | |
| **Yellow breast patch chroma (n. obs.: Corsica=3890; D-Rouvière=2007)** | | | | | | | | |
| **Fixed effects** | Est | SE | F | P | Est | SE | F | P |
| Year | **-0.051** | **0.018** | **$F_{1,13.15}$=12.673** | **0.003** | -0.040 | 0.021 | $F_{1,13.07}$=3.412 | 0.087 |
| Sex (fem) | **-0.068** | **0.004** | **$F_{1,2074.8}$=277.11** | **<0.001** | **-0.012** | **0.006** | **$F_{1,1015.5}$=4.100** | **0.043** |
| Year*sex(fem) | **0.020** | **0.004** | **$F_{1,2731.1}$=25.071** | **<0.001** | 0.001 | 0.006 | $F_{1,1248.6}$=0.101 | 0.750 |
| **Random effects** | Variance | Std. Dev. | | | Variance | Std. Dev. | | |
| Individual | 0.001 | 0.041 | | | 0.003 | 0.056 | | |
| Year | 0.002 | 0.045 | | | 0.007 | 0.086 | | |
| Site | 0.001 | 0.039 | | | - | - | | |
| Residual | 0.013 | 0.115 | | | 0.013 | 0.117 | | |
| | $R^2_{cond}$: 0.385; $R^2_{mar}$: 0.138 | | | | $R^2_{cond}$: 0.472; $R^2_{mar}$: 0.062 | | | |
| **Yellow breast patch brightness (n. obs.: Corsica=3890; D-Rouvière=2007)** | | | | | | | | |
| **Fixed effects** | Est | SE | F | P | Est | SE | F | P |
| Year | **-1.294** | **0.310** | **$F_{1,13.11}$=21.860** | **<0.001** | **-1.481** | **0.480** | **$F_{1,13.09}$=9.706** | **0.008** |
| Sex(fem) | **0.627** | **0.090** | **$F_{1,1904.7}$=48.007** | **<0.001** | 0.503 | 0.127 | **$F_{1,935.33}$=15.581** | **<0.001** |
| Year*sex(fem) | **-0.286** | **0.089** | **$F_{1,2600.6}$=10.271** | **0.001** | -0.004 | 0.126 | $F_{1,1169.5}$=0.001 | 0.971 |
| **Random effects** | Variance | Std. Dev. | | | Variance | Std. Dev. | | |
| Individual | 0.626 | 0.791 | | | 0.826 | 0.908 | | |
| Year | 1.453 | 1.205 | | | 3.655 | 1.911 | | |
| Site | 0.034 | 0.186 | | | - | - | | |
| Residual | 6.706 | 2.589 | | | 6.526 | 2.554 | | |
| | $R^2_{cond}$: 0.391; $R^2_{mar}$: 0.199 | | | | $R^2_{cond}$:0.507; $R^2_{mar}$: 0.169 | | | |

Table S8.2. Associations between the four colour components and the average temperature (Avg. temp.) and average precipitation (Avg. prec.) during moulting in Corsica and D-Rouvière. Significant ($p<0.05$) variables are in bold. $R^2_{cond}$ represents the variance explained by both the fixed and random effects included in the model, $R^2_{mar}$ represents the variance explained by the fixed factors alone. Predictors were scaled to mean of zero and standard deviation of unity.

| | **Corsica** | | | | **D-Rouvière** | | | |
|---|---|---|---|---|---|---|---|---|
| **Blue crown UV chroma (n. obs.: Corsica=3867; D-Rouvière=2068)** | | | | | | | | |
| **Fixed Effects** | Est | SE | F | P | Est | SE | F | P |
| Avg. temp. | -0.012 | 0.006 | $F_{1,12.0}$=3.537 | 0.084 | -0.036 | 0.053 | $F_{1,10.98}$=0.306 | 0.590 |
| Avg. prec. | 0.007 | 0.005 | $F_{1,12.0}$=1.343 | 0.268 | -0.039 | 0.053 | $F_{1,10.98}$=0.493 | 0.496 |
| Sex(fem) | **-0.035** | **0.0008** | **$F_{1,2205.4}$=1609.5** | **<0.001** | **-0.038** | **0.001** | **$F_{1,1185.9}$=1030.5** | **<0.001** |
| Year | | | | | -0.010 | 0.005 | $F_{1,11.02}$=3.688 | 0.081 |
| Avg. temp.*sex(fem) | **0.001** | **0.0007** | **$F_{1,3704.5}$=4.927** | **0.026** | 0.001 | 0.001 | $F_{1,1986.0}$=1.696 | 0.192 |
| Avg. prec.*sex(fem) | **-0.002** | **0.0008** | **$F_{1,3838.0}$=10.46** | **0.001** | 0.0005 | 0.001 | $F_{1,2039.3}$=0.261 | 0.609 |
| **Random Effects** | Variance | Std. Dev | | | Variance | Std. Dev | | |
| Individual | 0.0001 | 0.012 | | | 0.0001 | 0.013 | | |
| Year | 0.0005 | 0.022 | | | 0.0003 | 0.019 | | |
| Site | 0.00008 | 0.009 | | | - | - | | |
| Residual | 0.0004 | 0.020 | | | 0.0004 | 0.020 | | |
| | $R^2_{cond}$: 0.750; $R^2_{mar}$: 0.308 | | | | $R^2_{cond}$: 0.709; $R^2_{mar}$: 0.338 | | | |
| **Blue crown brightness (n. obs.: Corsica=3867; D-Rouvière=2068)** | | | | | | | | |
| **Fixed Effects** | Est | SE | F | P | Est | SE | F | P |
| Avg. temp. | -0.765 | 0.661 | $F_{1,12.0}$=1.010 | 0.334 | 1.414 | 0.670 | $F_{1,10.97}$=4.811 | 0.051 |
| Avg. prec. | 0.039 | 0.615 | $F_{1,12.0}$=0.008 | 0.927 | 0.763 | 0.661 | $F_{1,10.95}$=1.260 | 0.285 |
| Sex(fem) | **-2.379** | **0.118** | $F_{1,2118.3}$=400.85 | <0.001 | **-2.325** | **0.189** | **$F_{1,1108.0}$=149.87** | **<0.001** |
| Year | | | | | **-2.015** | **0.666** | **$F_{1,10.99}$=9.144** | **0.011** |
| Avg. temp.*sex(fem) | 0.206 | 0.117 | $F_{1,3833.5}$=3.109 | 0.077 | 0.084 | 0.181 | $F_{1,2036.1}$=0.218 | 0.640 |
| Avg. prec.*sex(fem) | 0.035 | 0.118 | $F_{1,3789.7}$=0.089 | 0.765 | -0.057 | 0.183 | $F_{1,2045.9}$=0.098 | 0.753 |
| **Random Effects** | Variance | Std. Dev | | | Variance | Std. Dev | | |
| Individual | 1.235 | 1.111 | | | 2.502 | 1.582 | | |
| Year | 5.766 | 2.401 | | | 5.821 | 2.413 | | |
| Site | 0.054 | 0.233 | | | - | - | | |
| Residual | 11.203 | 3.347 | | | 13.890 | 3.727 | | |
| | $R^2_{cond}$:0.444; $R^2_{mar}$: 0.094 | | | | $R^2_{cond}$:0.511; $R^2_{mar}$: 0.218 | | | |
| **Yellow breast patch chroma (n. obs.: Corsica=3890; D-Rouvière=2007)** | | | | | | | | |
| **Fixed Effects** | Est | SE | F | P | Est | SE | F | P |
| Avg. temp. | **-0.038** | **0.013** | **$F_{1,12.0}$=5.532** | **0.036** | 0.052 | 0.022 | $F_{1,11.01}$=4.272 | 0.063 |
| Avg. prec. | 0.025 | 0.012 | $F_{1,12.1}$=1.918 | 0.191 | -0.011 | 0.021 | $F_{1,10.99}$=0.009 | 0.922 |
| Sex(fem) | **-0.068** | **0.004** | **$F_{1,2070.1}$=276.90** | **<0.001** | **-0.012** | **0.006** | **$F_{1,1022.9}$=3.979** | **0.046** |
| Year | | | | | **-0.055** | **0.021** | **$F_{1,11.07}$=6.725** | **0.024** |
| Avg. temp.*sex(fem) | **0.013** | **0.004** | **$F_{1,3850.4}$=10.590** | **0.001** | **-0.012** | **0.005** | **$F_{1,1964.2}$=4.398** | **0.036** |
| Avg. prec. *sex(fem) | **-0.015** | **0.004** | **$F_{1,3816.5}$=13.514** | **<0.001** | **0.018** | **0.005** | **$F_{1,1989.0}$=9.713** | **0.001** |
| **Random Effects** | Variance | Std. Dev | | | Variance | Std. Dev | | |
| Individual | 0.001 | 0.041 | | | 0.003 | 0.056 | | |
| Year | 0.002 | 0.049 | | | 0.006 | 0.079 | | |
| Site | 0.001 | 0.039 | | | - | - | | |
| Residual | 0.013 | 0.115 | | | 0.013 | 0.011 | | |
| | $R^2_{cond}$: 0.395; $R^2_{mar}$=0.133 | | | | $R^2_{cond}$: 0.492; $R^2_{mar}$=0.137 | | | |
| **Yellow breast patch brightness (n. obs.: Corsica=3890; D-Rouvière=2007)** | | | | | | | | |
| **Fixed Effects** | Est | SE | F | P | Est | SE | F | P |
| Avg. temp. | **-1.050** | **0.408** | **$F_{1,12.0}$=7.654** | **0.017** | 0.154 | 0.562 | $F_{1,11.03}$=0.055 | 0.817 |
| Avg. prec. | 0.310 | 0.380 | $F_{1, 12.1}$=1.389 | 0.261 | 0.577 | 0.545 | $F_{1,11.01}$=0.874 | 0.369 |
| Sex(fem) | **0.625** | **0.090** | **$F_{1,1898.2}$=47.904** | **<0.001** | **0.501** | **0.127** | **$F_{1,941.90}$=15.501** | **<0.001** |
| Year | | | | | **-1.426** | **0.536** | **$F_{1,11.05}$=7.062** | **0.022** |
| Avg. temp.*sex(fem) | -0.143 | 0.089 | $F_{1,3858.7}$=2.560 | 0.109 | -0.044 | 0.124 | $F_{1,1982.5}$=0.128 | 0.719 |
| Avg. prec.*sex(fem) | **0.269** | **0.090** | **$F_{1,3797.1}$=8.899** | **0.002** | -0.142 | 0.124 | $F_{1,1975.4}$=1.290 | 0.256 |
| **Random Effects** | Variance | Std. Dev | | | Variance | Std. Dev | | |
| Individual | 0.620 | 0.788 | | | 0.820 | 0.906 | | |
| Year | 2.185 | 1.478 | | | 4.008 | 2.002 | | |
| Site | 0.036 | 0.190 | | | - | - | | |
| Residual | 6.705 | 2.589 | | | 6.530 | 2.555 | | |
| | $R^2_{cond}$: 0.412; $R^2_{mar}$=0.163 | | | | $R^2_{cond}$:0.524; $R^2_{mar}$=0.172 | | | |

As using scaled predictors may make the interpretation of the estimates a bit more confusing, below we present the results of the temporal trends in the colours and the association between colour and climate without scaling the predictors.

Table S8.3. Linear temporal trends for blue crown and yellow breast patch colourations of the blue tits in Corsica and D-Rouvière. Significant ($p<0.05$) variables are in bold. $R^2_{cond}$ represents the variance explained by both the fixed and random effects included in the model, $R^2_{mar}$ represents the variance explained by the fixed factors alone. Here predictors were not scaled.

| | **Corsica** | | | | **D-Rouvière** | | | |
|---|---|---|---|---|---|---|---|---|
| **Blue crown UV chroma (n. obs.: Corsica=3867; D-Rouvière=2068)** | | | | | | | | |
| **Fixed effects** | Est | SE | F | P | Est | SE | F | P |
| Year | **-0.004** | **0.001** | **$F_{1,13.01}=12.524$** | **0.003** | **-0.002** | **0.001** | **$F_{1,13.03}=5.038$** | **0.042** |
| Sex(fem) | **-1.981** | **0.402** | **$F_{1,2790.05}=24.236$** | **<0.001** | **-1.133** | **0.552** | **$F_{1,1408.23}=4.207$** | **0.040** |
| Year*sex(fem) | **0.0009** | **0.0001** | **$F_{1,2789.89}=23.383$** | **<0.001** | **0.0005** | **0.0002** | **$F_{1,1407.46}=3.927$** | **0.047** |
| **Random effects** | Variance | Std. Dev. | | | Variance | Std. Dev. | | |
| Individual | 0.0001 | 0.012 | | | 0.0001 | 0.012 | | |
| Year | 0.0003 | 0.019 | | | 0.0003 | 0.018 | | |
| Site | 0.00008 | 0.009 | | | - | - | | |
| Residual | 0.0004 | 0.020 | | | 0.0004 | 0.020 | | |
| | $R^2_{cond}$:0.740; $R^2_{mar}$: 0.365 | | | | $R^2_{cond}$:0.696; $R^2_{mar}$: 0.338 | | | |
| **Blue crown brightness (n. obs.: Corsica=3867; D-Rouvière=2068)** | | | | | | | | |
| **Fixed effects** | Est | SE | F | P | Est | SE | F | P |
| Year | -0.179 | 0.139 | $F_{1,13.05}=1.287$ | 0.276 | **-0.367** | **0.165** | **$F_{1,13.01}=5.670$** | **0.033** |
| Sex (fem) | -93.442 | 55.898 | $F_{1,2772.64}=2.794$ | 0.094 | 90.516 | 88.769 | $F_{1,1338.5}=1.039$ | 0.308 |
| Year*sex(fem) | 0.045 | 0.027 | $F_{1,2772.38}=2.653$ | 0.103 | -0.046 | 0.044 | $F_{1,1337.5}=1.093$ | 0.295 |
| **Random effects** | Variance | Std. Dev. | | | Variance | Std. Dev. | | |
| Individual | 1.261 | 1.123 | | | 2.487 | 1.577 | | |
| Year | 5.294 | 2.300 | | | 7.383 | 2.717 | | |
| Site | 0.055 | 0.234 | | | - | - | | |
| Residual | 11.178 | 3.343 | | | 13.890 | 3.727 | | |
| | $R^2_{cond}$:0.431; $R^2_{mar}$: 0.094 | | | | $R^2_{cond}$: 0.502; $R^2_{mar}$: 0.148 | | | |
| **Yellow breast patch chroma (n. obs.: Corsica=3890; D-Rouvière=2007)** | | | | | | | | |
| **Fixed effects** | Est | SE | F | P | Est | SE | F | P |
| Year | **-0.012** | **0.002** | **$F_{1,13.15}=12.673$** | **0.003** | -0.009 | 0.005 | $F_{1,13.07}=3.412$ | 0.087 |
| Sex (fem) | **-9.766** | **1.937** | **$F_{1,2735.02}=25.429$** | **<0.001** | -0.973 | 3.014 | $F_{1,1249.74}=0.104$ | 0.746 |
| Year*sex(fem) | **0.004** | **0.0009** | **$F_{1,2734.74}=25.071$** | **<0.001** | 0.0004 | 0.001 | $F_{1,1248.89}=0.101$ | 0.750 |
| **Random effects** | Variance | Std. Dev. | | | Variance | Std. Dev. | | |
| Individual | 0.001 | 0.041 | | | 0.003 | 0.056 | | |
| Year | 0.002 | 0.045 | | | 0.007 | 0.086 | | |
| Site | 0.001 | 0.039 | | | - | - | | |
| Residual | 0.013 | 0.115 | | | 0.013 | 0.117 | | |
| | $R^2_{cond}$: 0.385; $R^2_{mar}$: 0.138 | | | | $R^2_{cond}$: 0.472; $R^2_{mar}$: 0.062 | | | |
| **Yellow breast patch brightness (n. obs.: Corsica=3890; D-Rouvière=2007)** | | | | | | | | |
| **Fixed effects** | Est | SE | F | P | Est | SE | F | P |
| Year | **-0.306** | **0.073** | **$F_{1,13.11}=21.857$** | **<0.001** | **-0.357** | **0.116** | **$F_{1,13.09}=9.706$** | **0.008** |
| Sex(fem) | **137.099** | **42.582** | **$F_{1,2606.06}=10.366$** | **0.001** | 2.732 | 61.400 | $F_{1,1170.11}=0.002$ | 0.964 |
| Year*sex(fem) | **-0.067** | **0.021** | **$F_{1,2605.73}=10.271$** | **0.001** | -0.001 | 0.030 | $F_{1,1169.08}=0.001$ | 0.971 |
| **Random effects** | Variance | Std. Dev. | | | Variance | Std. Dev. | | |
| Individual | 0.626 | 0.791 | | | 0.826 | 0.908 | | |
| Year | 1.453 | 1.205 | | | 3.655 | 1.911 | | |
| Site | 0.034 | 0.186 | | | - | - | | |
| Residual | 6.706 | 2.589 | | | 6.526 | 2.554 | | |
| | $R^2_{cond}$: 0.391; $R^2_{mar}$: 0.199 | | | | $R^2_{cond}$:0.507; $R^2_{mar}$: 0.169 | | | |

Table S8.4. Associations between the four colour components and the average temperature (Avg. temp.) and average precipitation (Avg. prec.) during moulting in Corsica and D-Rouvière. Significant ($p<0.05$) variables are in bold. $R^2_{cond}$ represents the variance explained by both the fixed and random effects included in the model, $R^2_{mar}$ represents the variance explained by the fixed factors alone. Predictors were not scaled.

| | **Corsica** | | | | **D-Rouvière** | | | |
|---|---|---|---|---|---|---|---|---|
| **Blue crown UV chroma (n. obs.: Corsica=3867; D-Rouvière=2068)** | | | | | | | | |
| **Fixed Effects** | Est | SE | F | P | Est | SE | F | P |
| Avg. temp. | -0.023 | 0.011 | $F_{1,12.0}$=3.537 | 0.084 | -0.006 | 0.008 | $F_{1,10.98}$=0.306 | 0.590 |
| Avg. prec. | 0.018 | 0.013 | $F_{1,12.0}$=1.343 | 0.268 | -0.004 | 0.005 | $F_{1,10.97}$=0.493 | 0.496 |
| Sex(fem) | **-0.107** | **0.035** | **$F_{1,3673.2}$=9.370** | **0.002** | **-0.089** | **0.038** | **$F_{1,1994,25}$=5.276** | **0.021** |
| Year | | | | | **-0.002** | **0.001** | **$F_{1,11.0}$=3.687** | **0.081** |
| Avg. temp.*sex(fem) | **0.003** | **0.001** | **$F_{1,3704.5}$=4.927** | **0.026** | 0.002 | 0.001 | $F_{1,1986.02}$=1.696 | 0.192 |
| Avg. prec.*sex(fem) | **-0.006** | **0.001** | **$F_{1,3838.0}$=10.460** | **0.001** | 0.0006 | 0.001 | $F_{1,2039.37}$=0.261 | 0.609 |
| **Random Effects** | Variance | Std. Dev | | | Variance | Std. Dev | | |
| Individual | 0.0001 | 0.012 | | | 0.0001 | 0.013 | | |
| Year | 0.0005 | 0.022 | | | 0.0003 | 0.019 | | |
| Site | 0.00008 | 0.009 | | | - | - | | |
| Residual | 0.0004 | 0.020 | | | 0.0004 | 0.020 | | |
| | $R^2_{cond}$: 0.750; $R^2_{mar}$: 0.308 | | | | $R^2_{cond}$: 0.709; $R^2_{mar}$: 0.338 | | | |
| **Blue crown brightness (n. obs.: Corsica=3867; D-Rouvière=2068)** | | | | | | | | |
| **Fixed Effects** | Est | SE | F | P | Est | SE | F | P |
| Avg. temp. | -1.444 | 1.248 | $F_{1,12.0}$=1.010 | 0.334 | 2.343 | 1.111 | $F_{1,10.97}$=4.811 | 0.051 |
| Avg. prec. | 0.092 | 1.451 | $F_{1,12.0}$=0.008 | 0.927 | 0.842 | 0.729 | $F_{1,10.95}$=1.260 | 0.285 |
| Sex(fem) | **-11.529** | **5.209** | **$F_{1,3824.6}$=4.898** | **0.026** | -5.208 | 6.547 | $F_{1,2039.60}$=0.633 | 0.426 |
| Year | | | | | **-0.475** | **0.157** | **$F_{1,10.99}$=9.144** | **0.011** |
| Avg. temp.*sex(fem) | 0.390 | 0.221 | $F_{1,3833.4}$=3.109 | 0.077 | 0.140 | 0.301 | $F_{1,2036.15}$=0.218 | 0.640 |
| Avg. prec.*sex(fem) | 0.083 | 0.278 | $F_{1,3789.7}$=0.089 | 0.765 | -0.063 | 0.202 | $F_{1,2045.90}$=0.098 | 0.753 |
| **Random Effects** | Variance | Std. Dev | | | Variance | Std. Dev | | |
| Individual | 1.235 | 1.111 | | | 2.502 | 1.582 | | |
| Year | 5.766 | 2.401 | | | 5.821 | 2.413 | | |
| Site | 0.054 | 0.233 | | | - | - | | |
| Residual | 11.203 | 3.347 | | | 13.890 | 3.727 | | |
| | $R^2_{cond}$:0.444; $R^2_{mar}$: 0.094 | | | | $R^2_{cond}$:0.511; $R^2_{mar}$: 0.218 | | | |
| **Yellow breast patch chroma (n. obs.: Corsica=3890; D-Rouvière=2007)** | | | | | | | | |
| **Fixed Effects** | Est | SE | F | P | Est | SE | F | P |
| Avg. temp. | **-0.073** | **0.026** | **$F_{1,12.0}$=5.532** | **0.036** | 0.085 | 0.036 | $F_{1,11.01}$=4.272 | 0.063 |
| Avg. prec. | 0.059 | 0.030 | $F_{1,12.1}$=1.918 | 0.191 | -0.012 | 0.024 | $F_{1,10.99}$=0.009 | 0.922 |
| Sex(fem) | **-0.620** | **0.180** | **$F_{1,3839.2}$=11.894** | **<0.001** | 0.378 | 0.209 | $F_{1,1969.03}$=3.285 | 0.070 |
| Year | | | | | **-0.013** | **0.005** | **$F_{1,11.06}$=6.742** | **0.024** |
| Avg. temp.*sex(fem) | **0.024** | **0.007** | **$F_{1,3850.3}$=10.590** | **0.001** | **-0.020** | **0.009** | **$F_{1,1964.21}$=4.398** | **0.036** |
| Avg. prec. *sex(fem) | **-0.035** | **0.009** | **$F_{1,3816.5}$=13.514** | **<0.001** | **0.020** | **0.006** | **$F_{1,1989.01}$=9.713** | **0.001** |
| **Random Effects** | Variance | Std. Dev | | | Variance | Std. Dev | | |
| Individual | 0.001 | 0.041 | | | 0.003 | 0.056 | | |
| Year | 0.002 | 0.049 | | | 0.006 | 0.079 | | |
| Site | 0.001 | 0.039 | | | - | - | | |
| Residual | 0.013 | 0.115 | | | 0.013 | 0.011 | | |
| | $R^2_{cond}$: 0.395; $R^2_{mar}$=0.133 | | | | $R^2_{cond}$: 0.492; $R^2_{mar}$=0.137 | | | |
| **Yellow breast patch brightness (n. obs.: Corsica=3890; D-Rouvière=2007)** | | | | | | | | |
| **Fixed Effects** | Est | SE | F | P | Est | SE | F | P |
| Avg. temp. | **-1.986** | **0.772** | **$F_{1,12.0}$=7.652** | **0.017** | 0.251 | 0.915 | $F_{1,11.03}$=0.055 | 0.817 |
| Avg. prec. | 0.732 | 0.898 | $F_{1,12.1}$=1.389 | 0.261 | 0.636 | 0.601 | $F_{1,11.01}$=0.874 | 0.369 |
| Sex(fem) | 6.460 | 3.998 | $F_{1,3849.5}$=2.611 | 0.106 | 2.353 | 4.396 | $F_{1,1984.81}$=0.286 | 0.592 |
| Year | | | | | **-0.345** | **0.129** | **$F_{1,11.05}$=7.061** | **0.022** |
| Avg. temp.*sex(fem) | -0.271 | 0.169 | $F_{1,3858.7}$=2.560 | 0.109 | -0.072 | 0.202 | $F_{1,1982.57}$=0.128 | 0.719 |
| Avg. prec.*sex(fem) | **0.636** | **0.213** | **$F_{1,3797.1}$=8.899** | **0.002** | -0.156 | 0.138 | $F_{1,1975.45}$=1.290 | 0.256 |
| **Random Effects** | Variance | Std. Dev | | | Variance | Std. Dev | | |
| Individual | 0.620 | 0.788 | | | 0.820 | 0.906 | | |
| Year | 2.185 | 1.478 | | | 4.008 | 2.002 | | |
| Site | 0.036 | 0.190 | | | - | - | | |
| Residual | 6.705 | 2.589 | | | 6.530 | 2.555 | | |
| | $R^2_{cond}$: 0.412; $R^2_{mar}$=0.163 | | | | $R^2_{cond}$:0.524; $R^2_{mar}$=0.172 | | | |

**SM-9: Associations between blue crown brightness and yellow breast patch chroma in Rouviere**

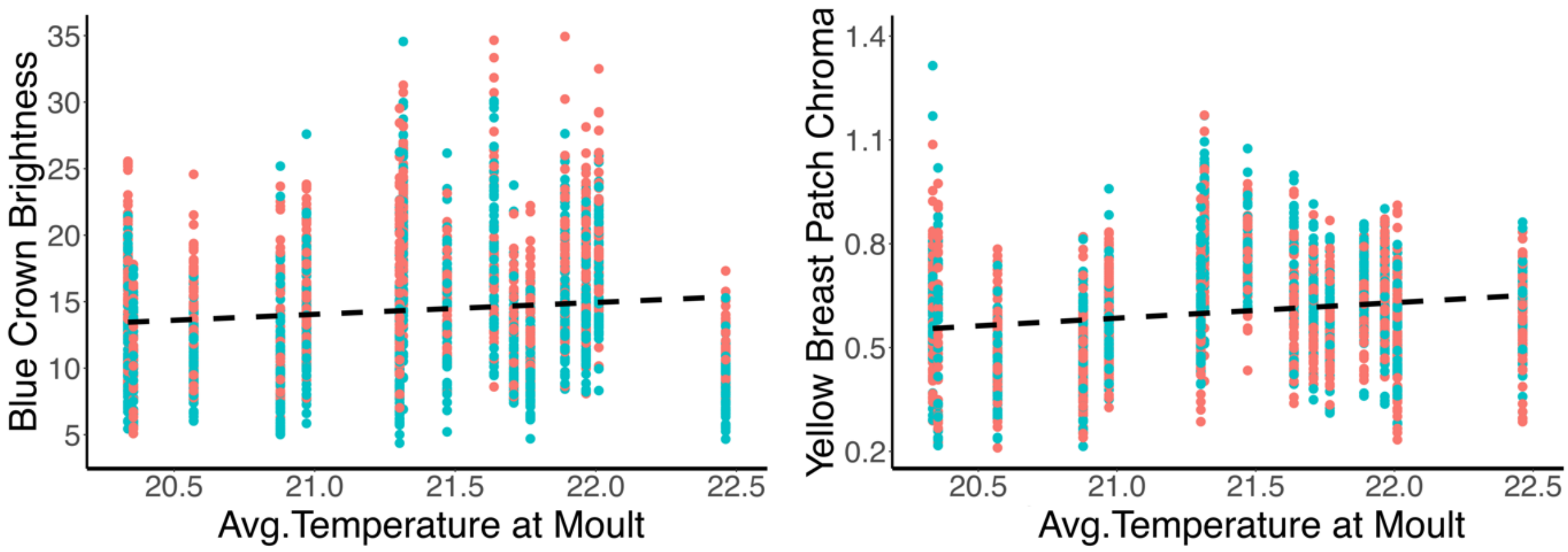

Fig. S9.1: In Rouviere, blue crown brightness (A) and yellow breast patch chroma (B) presented a positive marginally significant association with average temperature at moult. Blue and red dots represent males and females respectively. Dashed lines represent marginally significant results.

**SM-10: Permanent environment, year and residual variance of the animal models.**

Table S10.1: Additive genetic variance ($V_A$) estimates and 95% Credible Intervals (C.I.) obtained from the animal models of each coloured trait in each sex and population.

| | Corsica | | | | |
|---|---|---|---|---|---|
| | Males | | | Females | |
| | VA | 95% C.I. | | VA | 95% C.I. |
| Blue crown UV chroma | 0.0001 | [0.0001, 0.0002] | | 0.0001 | [0.0001, 0.0002] |
| Blue crown brightness | 0.647 | [0.00003, 1.503] | | 0.865 | [0.0004, 1.587] |
| Yellow breast patch chroma | 0.001 | [0.000000003, 0.002] | | 0.0004 | [0.0000000003, 0.012] |
| Yellow breast patch brightness | 0.701 | [0.051, 1.265] | | 0.328 | [0.0000004, 0.766] |
| | D-Rouvière | | | | |
| | Males | | | Females | |
| Blue crown UV chroma | 0.00006 | [0.00000000005, 0.0001] | | 0.0001 | [0.00006, 0.0002] |
| Blue crown brightness | 1.736 | [0.00000005, 3.273] | | 1.869 | [0.00000008, 3.247] |
| Yellow breast patch chroma | 0.003 | [0.0004, 0.0051] | | 0.0019 | [0.0000002, 0.003] |
| Yellow breast patch brightness | 0.388 | [0.00001, 0.947] | | 0.397 | [0.000001, 1.215] |

Table S10.2: Permanent environment variance ($V_{PE}$) estimates and 95% Credible Intervals (C.I.) obtained from the animal models of each coloured trait in each sex and population.

| | Corsica | | | | |
|---|---|---|---|---|---|
| | Males | | | Females | |
| | P.E. | 95% C.I. | | P.E. | 95% C.I. |
| Blue crown UV chroma | 0.000006 | [0.00000000000001, 0.001] | | 0.00001 | [0.00000000009, 0.00004] |
| Blue crown brightness | 0.581 | [0.00002, 1.546] | | 0.447 | [0.0001, 1.289] |
| Yellow breast patch chroma | 0.0006 | [0.0000000001, 0.001] | | 0.0008 | [0.000000008, 0.001] |
| Yellow breast patch brightness | 0.250 | [0.0000005, 0.799] | | 0.209 | [0.0000006, 0.655] |
| | D-Rouvière | | | | |
| | Males | | | Females | |
| Blue crown UV chroma | 0.00007 | [0.000000002, 0.0001] | | 0.0005 | [0.0000000002, 0.0001] |
| Blue crown brightness | 1.270 | [0.000001, 3.036] | | 1.270 | [0.000001, 3.036] |
| Yellow breast patch chroma | 0.0010 | [0.000000002, 0.003] | | 0.0009 | [0.000000005, 0.0026] |
| Yellow breast patch brightness | 0.350 | [0.000007, 0.982] | | 0.778 | [0.00001, 1.755] |

Table S10.3: Year variance ($V_{YR}$) and 95% Credible Intervals (C.I.) obtained from the animal models of each coloured trait in each sex and population.

| | Corsica | | | | |
|---|---|---|---|---|---|
| | Males | | | Females | |
| | YR | 95% C.I. | | YR | 95% C.I. |
| Blue crown UV chroma | 0.0005 | [0.0001, 0.001] | | 0.0004 | [0.00017, 0.0008] |
| Blue crown brightness | 7.953 | [2.838, 15.670] | | 5.956 | [1.934, 11.503] |
| Yellow breast patch chroma | 0.002 | [0.0009, 0.005] | | 0.002 | [0.0008, 0.005] |
| Yellow breast patch brightness | 2.028 | [0.673, 3.938] | | 1.912 | [0.589, 3.477] |
| | D-Rouvière | | | | |
| | Males | | | Females | |
| Blue crown UV chroma | 0.0005 | [0.0001, 0.011] | | 0.0003 | [0.0001, 0.0006] |
| Blue crown brightness | 9.383 | [2.571, 18.004] | | 10.379 | [3.710, 21.521] |
| Yellow breast patch chroma | 0.012 | [0.004, 0.025] | | 0.007 | [0.0025, 0.014] |
| Yellow breast patch brightness | 4.253 | [1.405, 8.443] | | 5.130 | [1.489, 9.904] |

Table S10.4: Residual variance ($V_{RES}$) and 95% Credible Intervals (C.I.) obtained from the animal models of each coloured trait in each sex and population.

| | Corsica | | | | |
|---|---|---|---|---|---|
| | Males | | | Females | |
| | Residual | 95% C.I. | | Residual | 95% C.I. |
| Blue crown UV chroma | 0.0004 | [0.0003, 0.0004] | | 0.0004 | [0.0003, 0.0004] |
| Blue crown brightness | 13.221 | [12.009, 14.339] | | 9.205 | [8.411, 9.992] |
| Yellow breast patch chroma | 0.013 | [0.012, 0.015] | | 0.012 | [0.011, 0.013] |
| Yellow breast patch brightness | 6.022 | [5.468, 6.626] | | 7.130 | [6.552, 7.706] |
| | D-Rouvière | | | | |
| | Males | | | Females | |
| Blue crown UV chroma | 0.0004 | [0.0003, 0.0005] | | 0.0004 | [0.0003, 0.0004] |
| Blue crown brightness | 13.758 | [11.938, 15.657] | | 13.633 | [3.710, 21.521] |
| Yellow breast patch chroma | 0.012 | [0.010, 0.013] | | 0.014 | [0.012, 0.016] |
| Yellow breast patch brightness | 5.223 | [4.634, 5.986] | | 7.572 | [6.656, 8.578] |

Table S10.5: Individual repeatability ($V_{PE}/V_P$) and 95% Credible Intervals (CI) obtained from the animal models of each coloured trait in each sex and population.

| | Corsica | | | | |
|---|---|---|---|---|---|
| | Males | | | Females | |
| | P.E. | 95% C.I. | | P.E. | 95% C.I. |
| Blue crown UV chroma | 0.006 | [<0.001, 0.024] | | 0.013 | [0.001, 0.047] |
| Blue crown brightness | 0.026 | [<0.001, 0.075] | | 0.027 | [<0.001, 0.082] |
| Yellow breast patch chroma | 0.035 | [<0.001, 0.106] | | 0.051 | [<0.001, 0.108] |
| Yellow breast patch brightness | 0.028 | [<0.001, 0.089] | | 0.022 | [<0.001, 0.066] |
| | D-Rouvière | | | | |
| | Males | | | Females | |
| Blue crown UV chroma | 0.066 | [<0.001, 0.139] | | 0.039 | [<0.001, 0.123] |
| Blue crown brightness | 0.049 | [<0.001, 0.122] | | 0.016 | [<0.001, 0.064] |
| Yellow breast patch chroma | 0.038 | [<0.001, 0.113] | | 0.036 | [<0.001, 0.109] |
| Yellow breast patch brightness | 0.035 | [<0.001, 0.102] | | 0.057 | [<0.001, 0.130] |

Table S10.6: Year explained variance ($V_{YR}/V_P$) and 95% Credible Intervals (CI) obtained from the animal models of each coloured trait in each sex and population.

| | Corsica | | | | |
|---|---|---|---|---|---|
| | Males | | | Females | |
| | Year | 95% C.I. | | Year | 95% C.I. |
| Blue crown UV chroma | 0.438 | [0.249, 0.647] | | 0.410 | [0.235, 0.599] |
| Blue crown brightness | 0.337 | [0.173, 0.530] | | 0.345 | [0.184, 0.551] |
| Yellow breast patch chroma | 0.150 | [0.058, 0.268] | | 0.165 | [0.065, 0.289] |
| Yellow breast patch brightness | 0.217 | [0.089, 0.360] | | 0.192 | [0.081, 0.318] |
| | D-Rouvière | | | | |
| | Males | | | Females | |
| Blue crown UV chroma | 0.460 | [0.275, 0.684] | | 0.341 | [0.177, 0.533] |
| Blue crown brightness | 0.341 | [0.188, 0.554] | | 0.373 | [0.185, 0.570] |
| Yellow breast patch chroma | 0.413 | [0.235, 0.617] | | 0.285 | [0.133, 0.462] |
| Yellow breast patch brightness | 0.396 | [0.225, 0.613] | | 0.352 | [0.176, 0.544] |

## SM-11: TEMPORAL TRENDS IN THE BREEDING VALUES

ROUVIERE

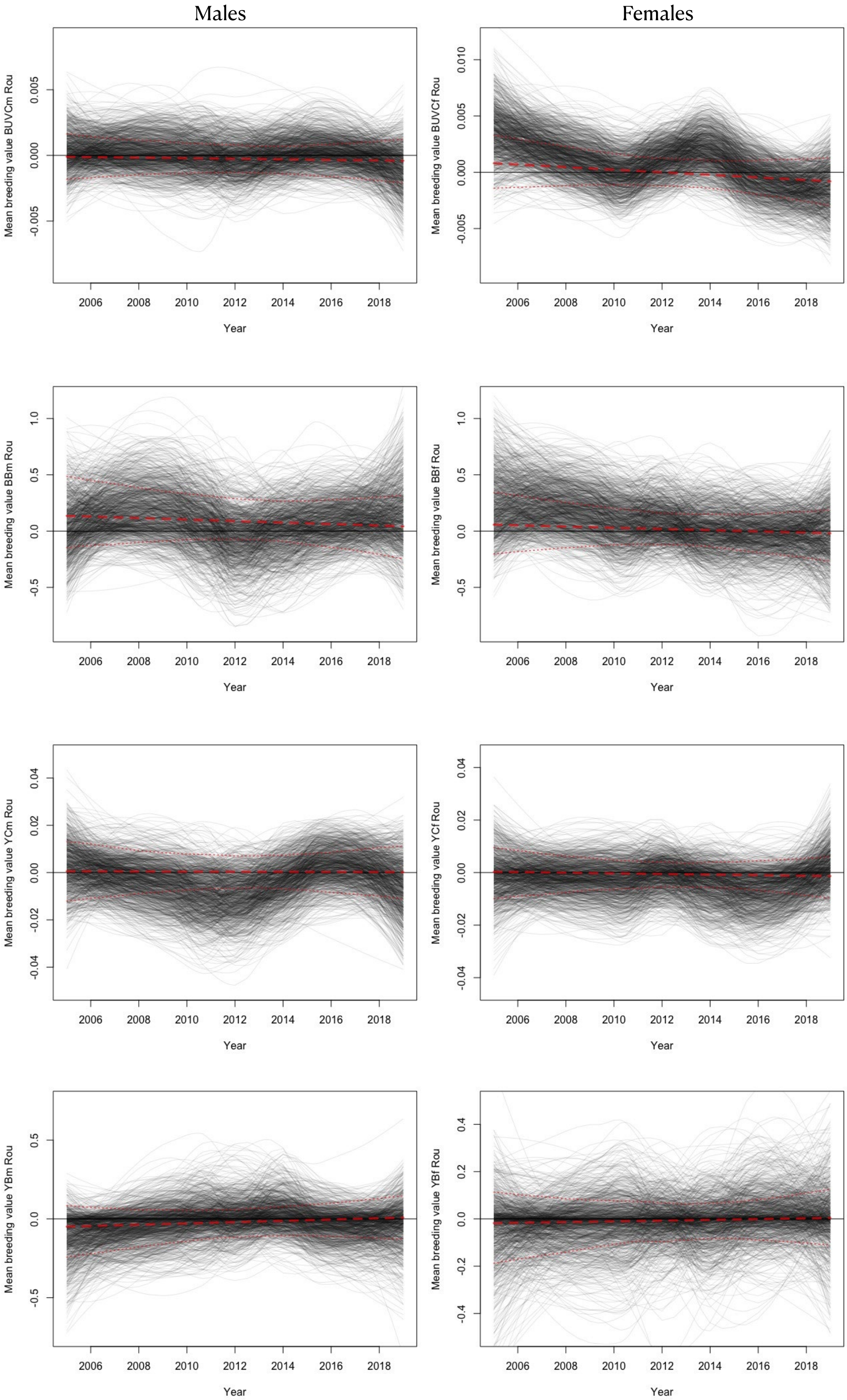

Figure S11.1: Temporal trends in the breeding values of the coloured traits in D-Rouvière. Dashed red lines represent the linear regression of the predicted breeding values on the mean year an individual was present in the population (see methods), thick lines represent the posterior mode of the regression and thin lines the 95% C.I.. Black lines represent the change in the breeding values allowing for a nonlinear change using splines, and represent a sample

from the 1000 different MCMC iterations. This graph was made following the code provided in Bonnet et al. (2019).

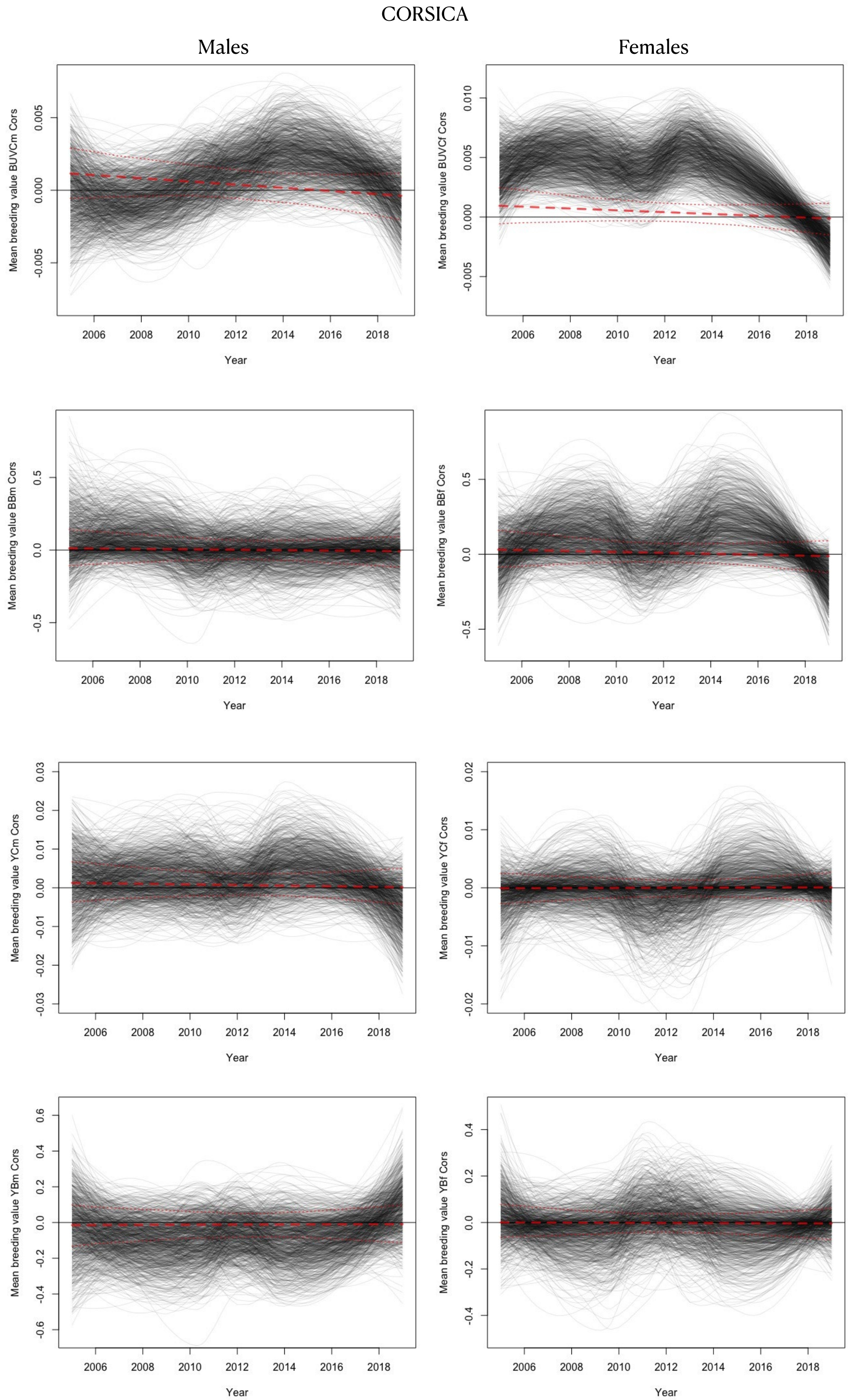

Figure S11.2: Temporal trends in the breeding values of the coloured traits in Corsica. Dashed red lines represent the linear regression of the predicted breeding values on the mean year an individual was present in the population (see methods), thick lines represent the posterior mode of the regression and thin lines the 95% C.I.. Black lines represent the change in the breeding values allowing for a nonlinear change using splines, and represent a sample from the 1000 different MCMC iterations. This graph was made following the code provided in Bonnet et al. (2019).

## SM-12: Reduction in the expression of the studied colourations

Table S12.1: Reduction in the coloured traits between the beginning (period 2005-2009) and the end (period 2015-2019) in Corsica and D-Rouvière.

| Corsica | | | | | |
|---|---|---|---|---|---|
| | Avg. 2005-2009 | SE 2005-2009 | Avg. 2015-2019 | SE 2005-2009 | Change |
| Blue UV Chroma | 0.3727 | 0.001 | 0.3233 | 0.0005 | -13.25% |
| Blue Brightness | 13.586 | 0.118 | 11.922 | 0.107 | -12.24% |
| Yellow Chroma | 0.7963 | 0.004 | 0.6736 | 0.002 | -15.40% |
| Yellow Brightness | 16.674 | 0.092 | 13.166 | 0.065 | -21.03% |
| D-Rouviére | | | | | |
| | Avg. 2005-2009 | SE 2005-2009 | Avg. 2015-2019 | SE 2005-2009 | Change |
| Blue UV Chroma | 0.3599 | 0.001 | 0.3264 | 0.0009 | -9.30% |
| Blue Brightness | 16.347 | 0.215 | 12.455 | 0.155 | -23.80% |
| Yellow Chroma | 0.6517 | 0.008 | 0.5772 | 0.004 | -11.43% |
| Yellow Brightness | 17.023 | 0.163 | 14.010 | 0.094 | -17.69% |

## SM-13: Reduction in the coefficients of variation of the 4 colour components

We explored whether there was a reduction in the variability of our coloured traits with time. To do that we tested whether there has been a reduction in the coefficients of variation (C.V.) when comparing the first five years (beginning: 2005-2009) to the last five years (end: 2015-2019). We fitted a set if Linear Models (LMs), one for each trait and population, in which the yearly C.V. were included as dependent variables and the period (beginning vs end) was included as explanatory term. Overall, our results suggest that there has been a decrease in the variability of the studied traits (SM-10 Table S1, SM-10 Fig. S1).

Table S13.1: Differences between the coefficients of variation (C.V.) in the beginning (2005-2009) and in the end (2015-2019) of our study period in Corsica and D-Rouvière. Significant ($p<0.05$) variables are in bold.

| | *Corsica (n=10)* | | | | *D-Rouvière (n=10)* | | | |
|---|---|---|---|---|---|---|---|---|
| | Est | SE | F | P | Est | SE | F | P |
| | C.V. Blue Crown UV Chroma | | | | | | | |
| Period(*end*) | **-3.880** | **0.486** | **$F_{1,8}$=63.592** | **<0.001** | **-2.248** | **0.507** | **$F_{1,8}$=19.636** | **0.002** |
| | C.V. Blue Crown Brightness | | | | | | | |
| Period(*end*) | -1.451 | 0.897 | $F_{1,8}$=2.614 | 0.144 | -0.389 | 2.517 | $F_{1,8}$=0.024 | 0.880 |
| | C.V. Yellow Breast Patch Chroma | | | | | | | |
| Period(*end*) | **-2.070** | **0.815** | **$F_{1,8}$=6.450** | **0.034** | -3.960 | 4.259 | $F_{1,8}$=0.864 | 0.379 |
| | C.V. Yellow Breast Patch Brightness | | | | | | | |
| Period(*end*) | -1.726 | 0.870 | $F_{1,8}$=3.931 | 0.082 | -4.156 | 1.865 | $F_{1,8}$=4.964 | 0.056 |

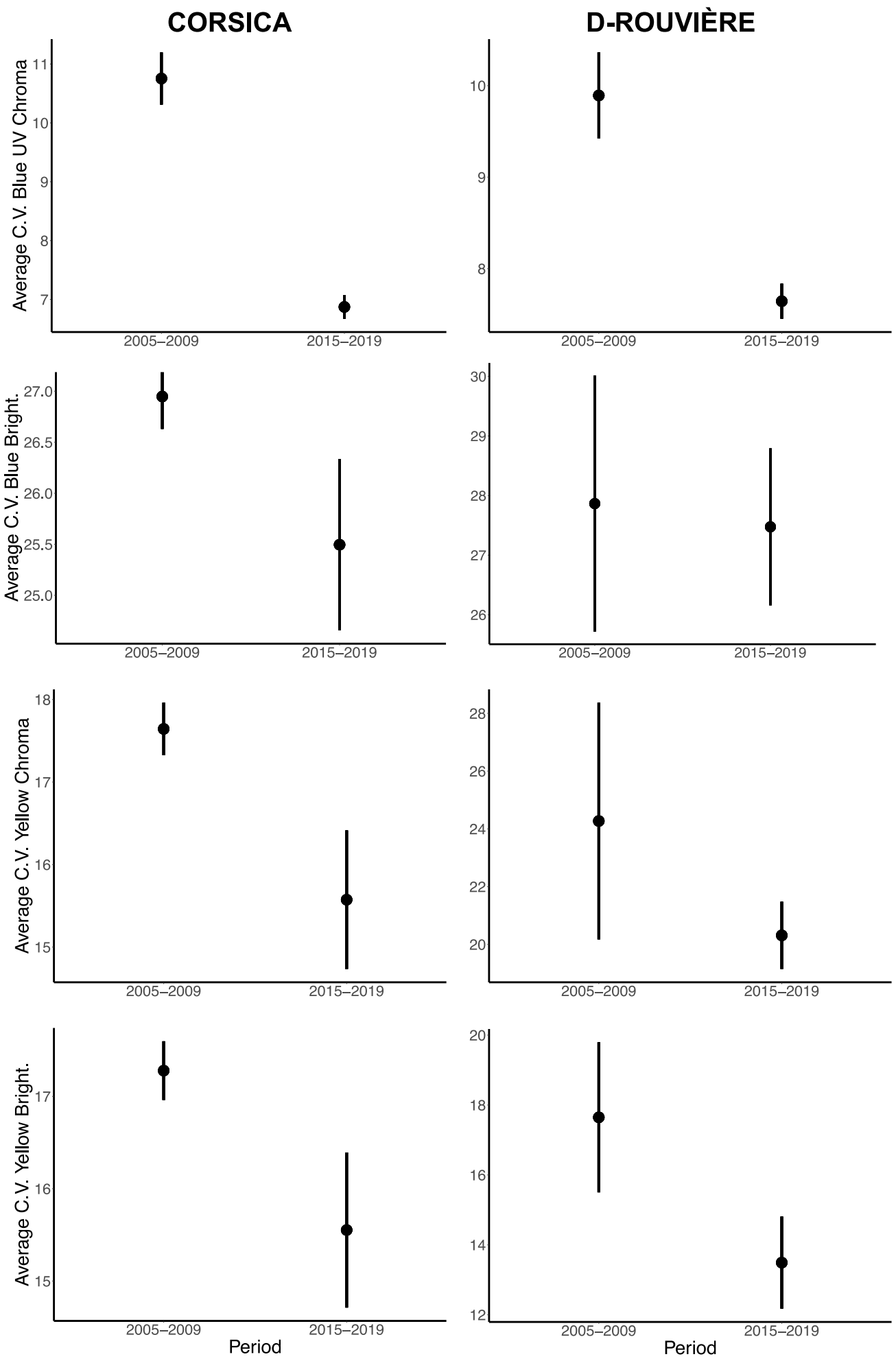

Fig. S13.1: Mean ± SE values of the coloured components coefficients of variation (C.V.) in the beginning (2005-2009) and in the end (2015-2019) of our study period in Corsica and D-Rouvière.